\documentclass[10pt,journal,final]{IEEEtran}

\usepackage[cmex10]{amsmath} 
\usepackage{amssymb,mathrsfs}
\usepackage{epsfig,epsf,subfigure,graphicx,graphics}
\usepackage{url}

\newtheorem{remark}{Remark}

\newcommand{\SNR}{\mathsf{SNR}}

\newcommand{\C}{\mathfrak{C}}

\newcommand{\aaaa}{\mathrm{(a)}}

\newcommand{\lp}{\left(}
\newcommand{\rp}{\right)}

\newcommand{\lbp}{\left\{}
\newcommand{\rbp}{\right\}}
\newcommand{\ul}{\underline}

\newcommand{\mcal}{\mathcal}

\newcommand{\msf}{\mathsf}

\newcommand{\wtild}{\widetilde}
\newcommand{\mb}{\mathbf}
\newcommand{\mbf}{\mathbf}
\newcommand{\mbb}{\mathbb}
 \DeclareMathOperator{\erfc}{erfc}

\graphicspath{{fig/}}

\title{Coding and System Design for Quantize-Map-and-Forward Relaying}
\author{Vinayak~Nagpal,~\IEEEmembership{Member,~IEEE,} I-Hsiang~Wang,~\IEEEmembership{Member,~IEEE,} Milos~Jorgovanovic,~\IEEEmembership{Student~Member,~IEEE,} David~Tse,~\IEEEmembership{Fellow,~IEEE} and Borivoje~Nikoli\'{c}~\IEEEmembership{Senior~Member,~IEEE}
\thanks{The material in this paper was presented in part at the Annual Allerton Conference on Communication, Control, and Computing, Monticello, Illinois, USA, September 2010.}
\thanks{At the time of submission, all authors were with the Department of EECS, University of California at Berkeley, Berkeley, California 94720 USA. e-mail: vinayak.nagpal@nokia.com, i-hsiang.wang@epfl.ch, \{milos,dtse,bora\}@eecs.berkeley.edu.}
}

\begin{document}
\maketitle


\begin{abstract}
In this paper we develop a low-complexity coding scheme and system design framework for the half duplex relay channel based on the Quantize-Map-and-Forward (QMF) relaying scheme. The proposed framework allows linear complexity operations at all network terminals.
We propose the use of binary LDPC codes for encoding at the source and LDGM codes for mapping at the relay. We express \emph{joint decoding} at the destination as a belief propagation algorithm over a factor graph. This graph has the LDPC and LDGM codes as subgraphs connected via probabilistic constraints that model the QMF relay operations. We show that this coding framework extends naturally to the high SNR regime using bit interleaved coded modulation (BICM).
We develop density evolution analysis tools for this factor graph and demonstrate the design of practical codes for the half-duplex relay channel that perform within $1$dB of information theoretic QMF threshold.
\end{abstract}

\section{Introduction}
Cooperative relaying has been proposed as a promising technique to resolve the increasing demand for data throughput in wireless networks. Recently a lot of progress has been made in establishing the theoretical foundations of cooperative communication. 
To apply these principles towards the design of practical wireless systems, various system design tradeoffs must be taken into consideration. 
This paper presents progress towards this goal.
We propose a system design and coding framework for quantize-map-and-forward (QMF) \cite{2009arXiv0906.5394A} relaying that has low complexity and performs close to information theoretic bounds. 

\subsection{Cooperative Systems}

A cooperative wireless link typically consists of an information source, a destination  and one or more cooperating \emph{half duplex} relays. The relays are usually assumed to operate \emph{in-band}. i.e. no additional channel resources are allocated for cooperation.
 Without loss of generality, it is assumed that relays use time-division-duplexing i.e. they listen to transmission from the source for some fraction of total time, then forward a description of their observation in the remaining fraction.

There are several aspects involved in the design of a cooperative relaying system. 
Listening fractions and forwarding schemes must be determined for each relay. 
Suitable modulation and channel coding schemes must be designed for various terminals. Rate adaptation mechanisms must be considered to account for changes in availability of relays and channel strengths. Practical constraints must be considered e.g. minimizing the overall system complexity, 
reuse of building blocks from traditional (non-cooperative) systems as much as possible, compatibility with protocols at higher layers and handling of system imperfections like synchronization, channel estimation errors etc. 
In this paper, we focus on the coding and signaling aspects of cooperative relaying. Other components are discussed in brief towards the end of the paper

\subsection{Relaying Schemes}
\begin{figure}[htbp]
{\centering
\includegraphics[width=2in]{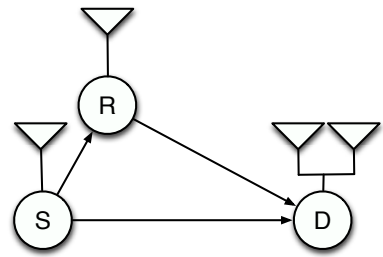}
\caption{Example relay network. With multiple antennas at destination, source-relay cooperation provides additional degrees of freedom for communication.}
\label{fig:example}
}
\end{figure}
Most wireless systems operate at moderate to high SNR i.e. in a regime where transmit power is not the major limiting factor for link capacity. At high SNR, the link capacity is limited by spatial degrees of freedom. Relay cooperation is of special interest for practical systems because it has the potential to provide additional spatial degrees of freedom. For illustration, consider a relay channel with single-antenna source, a half-duplex single antenna relay and a destination with two antennas shown in Fig.~\ref{fig:example}. Since the destination has multiple antennas, the source can spatially multiplex traffic to the destination using relay cooperation. If the source-to-relay channel is strong, this network can approach the high-SNR performance of the $2\times2$ MIMO channel \cite{5205885}. 


Various strategies for relay cooperation are proposed in 
literature. Among these amplify-and-forward (AF), decode-and-forward (DF) and compress-and-forward (CF) \cite{1056084,1362898} have received the most attention. Under DF, the relay decodes the source's message and forwards a hard estimate of it, whereas under AF and CF it forwards a soft estimate without explicitly decoding it. 
In DF and CF, the relay maps its estimate to a random codeword before forwarding, whereas in AF it forwards an uncoded signal. 
The QMF scheme \cite{2009arXiv0906.5394A} also uses soft estimate forwarding with random coding similar to CF. 
For the example network in Fig.~\ref{fig:example}, the CF and QMF schemes are close to optimal at high SNR. In fact they achieve within one bit/sec/Hz to the information-theoretic capacity \cite{4797531, 2009arXiv0906.5394A}. 
An intuitive explanation for why QMF/CF performs better than both AF and DF is given in the context
of the example in Fig.~\ref{fig:example} below.

In the example network of Fig.~\ref{fig:example}, the destination receives continuously from the source. The relay receives a stronger version, when it is listening. Since the destination has two antennas, it can resolve simultaneous transmissions from the source and relay. In order to achieve spatial multiplexing, the relay should extract the less significant bits (from its observation), which the destination cannot resolve, and forward them. Under AF, the relay only forwards the more significant bits, which the destination can already resolve. Therefore AF cooperation provides limited benefit. Under DF, the relay decodes the entire message before it forwards anything. Since the listening time is limited, this approach is inefficient. The QMF/CF schemes implicitly extract the less significant bits from the relay's observation by using quantization/compression and random mapping. Therefore these schemes provide the most cooperation gain. 

Despite having similar performance for the single relay network, the CF and QMF schemes have significant differences.  In the conventional CF scheme, the relay compresses its observed signal and performs a random code mapping before forwarding. The compression rate is chosen in order for the destination to perform \emph{two-step decoding} i.e. first decode compressed signals from relay and then use it as side information to decode the message from source. 
For configurations that involve multiple relays, two-step decoding is sub-optimal and conventional CF is not within bounded gap from information-theoretic capacity \cite{2009arXiv0906.5394A}.
Even for the single-relay configuration, conventional CF requires that the relay have full knowledge of the quality of its forward channel. This introduces a large estimation and feedback overhead for fading channels and
increases the complexity of rate adaptation schemes. 

Under QMF, the relay quantizes its received signal at noise level, randomly maps it to a codeword and forwards it. Unlike CF, the quantization and mapping is performed without regard to the quality of forward channel at the relay. This reduces the channel estimation and feedback overhead for the link. It also simplifies rate adaptation protocols. Additionally, QMF uses \emph{joint decoding} (as opposed to successive decoding) and performs within bounded gap from capacity for networks having an arbitrary number of relays \cite{2009arXiv0906.5394A}. QMF has played a key role in several recent information theoretic results on cooperative networks \cite{2010arXiv1002.3188L,4797531,5205885,WangTse_09}.
Due to these favorable properties, the QMF scheme is superior to CF from the perspective of practical cooperative systems.


Since mapping at a QMF relay is performed without any knowledge of forward channel strength, side information from relays cannot be decoded at the destination independently. QMF requires \emph{joint decoding} of the message (from source) and side information (from relays) \cite{2009arXiv0906.5394A}. This presents a unique challenge because joint decoding typically requires higher complexity and makes it harder to design a practical cooperative coding scheme. The key contribution of this paper is to develop a low-complexity cooperative coding framework for QMF that significantly reduces the complexity of joint decoding and yet performs close to information theoretic bounds. 
\subsection{Related Work}
Majority of previous work on code design for cooperative relaying is focused on the DF scheme. 
DF relays fully decode the source's message. Therefore, DF coding schemes involve partitioning a large codebook into two parts. The source transmits one part of the codeword and the relay transmits the remaining part \cite{1204784,1023492,1261324}. Turbo code designs which perform $\approx1$dB away from the DF information theoretic threshold are demonstrated in \cite{1532486, 4303346}. LDPC profiles are developed for DF in \cite{4107948}. A bilayer LDPC structure \cite{4305411} and the protograph method \cite{5513451} has been used to get LDPC designs  $\leq0.5$dB from the DF threshold. The bilayer structure is extended for use at high SNR using bit-interleaved coded modulation (BICM) \cite{4259765}. 
As for CF relaying, a coding scheme using a combination of LDPC and irregular repeat accumulate (IRA) codes is presented in \cite{1721040}. Rateless coding schemes are developed in \cite{5604326}. 
As for QMF relaying, a coding scheme is proposed in independent work \cite{2010arXiv1005.1284O} based on lattice strategies. The scheme in \cite{2010arXiv1005.1284O} reduces the complexity of mapping at the relay to polynomial-time while the joint decoding complexity remains exponential-time. 

\subsection{Summary of Results}
In this paper, a coding scheme for QMF relaying with linear complexity encoding at the source, mapping at the relays and joint decoding at the destination is developed. For a network with one relay, the proposed scheme performs within $(0.5-1)$dB gap from the information-theoretic QMF threshold. For the code design example considered in Sec.~\ref{sec_example}, the QMF threshold is $\approx1.5$dB better than DF. The key techniques used in this paper are summarized as follows:
\label{subsec_contributions}
\begin{IEEEenumerate}
\item \emph{BICM:} Design of \emph{binary} channel codes with standard higher order signal constellations is considered based on the widely used BICM technique \cite{669123}. 
\item \emph{LDPC-LDGM:} The scheme uses low density parity check (LDPC) codes at the source for channel coding and low density generator matrix (LDGM) codes at the relays for mapping. 
\item \emph{Joint Factor Graph:} The joint decoding procedure at the destination is formulated as a belief propagation algorithm over a factor graph. This graph contains the original channel code (LDPC) and relay mapping functions (LDGM) as subgraphs connected via probabilistic constraints that model the QMF relay operations.  
\item \emph{Practical Decoding Algorithm:} Using a DBLAST space-time architecture, scalar quantization procedure at relays and specific choice of component codes, the resulting factor graph is greatly simplified, making it suitable for practical decoder implementation.
\item \emph{Code Design:}  Density evolution analysis tools \cite{910578,910577} are developed for the systematic design of joint LDPC-LDGM factor graphs.
\end{IEEEenumerate}

\subsection{Organization}
In Sec.~\ref{sec_two}, the coding framework for QMF and corresponding joint decoding algorithm is developed. The treatment focusses on a canonical system model with one relay and binary inputs.
In Sec.~\ref{sec_densityevolution}, density evolution and code design tools are developed. In Sec.~\ref{subsec_BICM}, the framework is extended to the high SNR regime i.e. for high order modulation inputs using BICM. In Sec.~\ref{sec_example}, the design of codes for an example cooperative link is demonstrated. Finally in Sec.~\ref{sec_discussion} a sketch is provided for extending the proposed framework to scenarios with multiple relays.

\section{Coding Framework}
\label{sec_two}
\subsection{System Model} \label{sec_problem}
Initially, this paper focuses on the design of codes for a \emph{binary memoryless symmetric} (BMS) relay channel as described below. In Sec~\ref{subsec_BICM}, this model is extended to high order modulation inputs for use at high SNR. 

\begin{figure}[htbp]
{\centering
\includegraphics[width=3in]{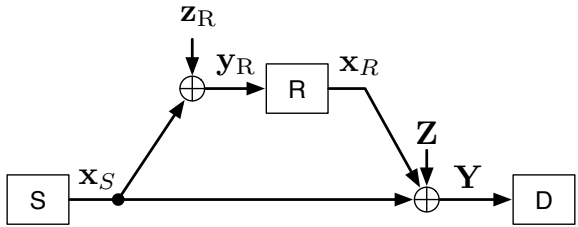}
\caption{Half-Duplex Binary Input Gaussian Relay Channel}
\label{fig_HDR}
}
\end{figure}

The BMS Gaussian relay channel has three half-duplex terminals: source ($S$), relay ($R$) and destination ($D$) with binary input additive white Gaussian (BIAWGN) channels between them, as shown in Fig.~\ref{fig_HDR}. $R$ listens for a fraction $f \in [0,1]$ of the total communication time and transmits for the fraction $(1-f)$. The block lengths for the transmitted codewords at $S$ and $R$ are $N_{S}$ and $N_{R}$ respectively. They satisfy the half-duplex constraint $N_{R} = (1-f) N_{S}$. 
The codeword messages sent from $S$ and $R$ are $\mathbf{b}_{S}\in\{0,1\}^{N_{S}}$ and  $\mathbf{b}_{R}\in \{0,1\}^{N_{R}}$ respectively. The corresponding transmitted signals are $\mbf{x}_{S}\in\{\pm \sqrt{P_{S}}\}^{N_{S}}$ and $\mbf{x}_{R}\in\{\pm \sqrt{P_{R}}\}^{N_{R}}$ where $P_{S}$ and $P_{R}$ are \emph{per-node symbol constraints} on average power i.e. $E|x_{S,i}^{2}|\leq P_{S}$ and $E|x_{R,i}^{2}|\leq P_{R}$. 
Bold-face lower case letters are used to denote a sequence of symbols. 

Multiple ($M$) receive antennas are assumed at the destination. This permits consideration of network scenarios where \emph{cooperative spatial multiplexing} is possible \cite{4698542}\cite{5205885}. An example scenario with $M=2$ is discussed in the Appendix.
The received signals at $D$ and $R$ are denoted as $\mbf{y}_{i}\in\mathbb{C}^{M}$ and $y_{R,j}$ for each symbol time $i\in\{1,2,\ldots,N_{S}\}$ and $j\in\{1,2,\ldots,fN_{S}\}$ respectively. They are modeled as follows:
\begin{align*}
\mbf{y}_{i} &= \mbf{h}_1x_{S,i} + \mbf{h}_2 x'_{R,i} + \mbf{z}_{i},\ 
y_{R,j} = h_{R}x_{S,j} + z_{R,j}
\end{align*} 

Here $\mbf{h}_1,\mbf{h}_2,h_{R}$ denote the corresponding channel gains. $x'_{R,i}=0$ and $\mbf{h}_{2}=\mbf{0}$ for $i\in\{1,2,\ldots,fN_{S}\}$ when $R$ is listening. For the remaining time $x'_{R,i}~=~x_{R,i-fN_{S}}$, $i\in\{fN_{S}+1,\ldots,N_{S}\}$. $\mathbf{z_{i}}$ and $z_{R,j}$ are i.i.d. zero-mean Gaussian noise vectors with identity covariance matrices. All the channel observations at $D$ are denoted by $\mbf{Y}\in\mathbb{C}^{ M\times N_{S}}$ i.e. $\mbf{Y}=[\mbf{y}_{1}\ \mbf{y}_{2}\ \ldots \ \mbf{y}_{N_{S}}]$. Observations at $R$ are denoted as $\mbf{y}_{R} \in \mathbb{C}^{1\times fN_{S}}$ i.e. $\mbf{y}_{R}=[y_{R,1}\ \ldots \ y_{R,fN_{S}}]$. 
The channel is characterized by the following parameters: $\SNR_{SR}=P_{S}|{h}_{R}|^{2}$, $\SNR_{SD}=P_{S}||\mbf{h}_{1}||^{2}$ and $\SNR_{RD}=P_{R}||\mbf{h}_{2}||^{2}$.


\subsection{Quantize-Map-Forward Scheme}
The quantize-map-and-forward scheme \cite{2009arXiv0906.5394A} is summarized as follows. $S$ has a sequence of messages $m_k
\in\{1,\ldots,2^{N_{S}\mcal{R}}\}$, $k=1,2,\ldots$ to be transmitted. At
both $S$ and $R$, codebooks $\mcal{C}_{S}$ and $\mcal{C}_{R}$ are created respectively.
$S$ maps each message to one of its codewords
and transmits it using $N_{S}$ symbols resulting in an overall
transmission rate of $\mcal{R}$. Relay listens to the first $fN_{S}$ time symbols of each block. It quantizes its observation at noise level i.e. the quantization distortion is equal to the noise power at the relay. 
Relay maps the quantized bits to a codeword in $\mcal{C}_{R}$. It transmits this codeword using $(1-f)N_{S}$ symbols. The destination $D$ attempts to decode the message sent by $S$ from received signals $(\mbf{Y})$. In order to decode, $D$ must know all channel parameters $\SNR_{SD},\SNR_{RD}$ and $\SNR_{SR}$, the relay listening fraction $f$ and both codebooks $\mcal{C}_{S}$ and $\mcal{C}_{R}$.

It is assumed that $\SNR_{SD}$,$\SNR_{RD}$ are measured at $D$ and $\SNR_{SR}$ is measured at $R$ using pilot symbols. It is further assumed that $\SNR_{SR}$ is forwarded to $D$ by $R$. The estimation and forwarding overhead of these steps is ignored for the analysis presented in this paper.  

\subsection{Factor Graph for Joint Decoding}
\label{sec_coding}
In the context of the system model and cooperation scheme outlined above, let us focus on binary \emph{linear} codebooks $\mathcal{C}_{S}^{b}$ and $\mathcal{C}_{R}^{b}$.
These can be represented as bipartite Tanner graphs using respective parity check matrices. In such a representation bit (variable) nodes represent the codeword and check (function) nodes represent parity constraints that must be satisfied in order for the codeword to be valid. 
Let us consider the maximum {\it a posteriori} (MAP) rule for joint decoding at $D$. In this subsection, joint decoding is expressed as a sum-product algorithm over a factor graph that contains the Tanner graphs of component codes ($\mathcal{C}_{S}^{b},\mathcal{C}_{R}^{b}$) as sub-graphs connected via probabilistic constraints that represent the QMF relaying operation \cite{5706940}\cite{910572}\cite{825794}. 

Joint decoding involves searching for the codeword $\mathbf{b}_{S}\in \mcal{C}_{S}^{b}$ that maximizes the {\it a posteriori} probability $p\lp \mb{b}_{S} | \mbf{Y}\rp$. An efficient way to do this search is to consider the bitwise maximum {\it a posteriori} (MAP) decoder, where the aim is to compute $p\lp b_{S,i} | \mbf{Y}\rp~=~\sum_{\sim b_{S,i}} p\lp \mb{b}_{S} | \mbf{Y}\rp
$ for all $i=1,2,\ldots N_{S}$.
\begin{align*}
p\lp \mb{b}_{S} | \mb{Y}\rp &= \sum_{\mb{b}_{R}} \frac{f\lp \mb{Y}| \mb{b}_{S}, \mb{b}_{R} \rp p\lp \mb{b}_{S}, \mb{b}_{R}\rp}{f\lp \mb{Y}\rp} \\
&\propto \sum_{\mb{b}_{R}} f\lp \mb{Y}| \mb{b}_{S}, \mb{b}_{R} \rp p\lp \mb{b}_{S}, \mb{b}_{R}\rp.
\end{align*}
For the first $fN_{S}$ bits, $R$ is listening and $D$ observes an interference-free signal from $S$. During the remaining transmissions, $D$ observes a superposition of signals from $S$ and $R$. Therefore, the first term $f\lp \mb{Y}| \mb{b}_{S}, \mb{b}_{R} \rp$ factorizes as follows:
\begin{align*}
&f(\mb{Y}| \mb{b}_{S}, \mb{b}_{R}) \\* &= \prod_{i=1}^{f N_{S}} f(\mbf{y}_{i}|b_{S,i}) \prod_{j=1}^{N_{R}} f(\mbf{y}_{(f N_{S}+j)}|b_{{S},(f N_{S}+j)},b_{R,j})
\end{align*} 
The codes $\mathcal{C}_{S}^{b}$ and $\mathcal{C}_{R}^{b}$ have characteristic functions $\mbf{1}(\mathbf{b}_{S}\in \mathcal{C}_{S}^{b})$ and $\mbf{1}(\mathbf{b}_{R}\in \mathcal{C}_{R}^{b})$ respectively. 
\begin{align*}
p\lp \mb{b}_{S}, \mb{b}_{R}\rp &= p\lp \mb{b}_{S} \rp p\lp \mb{b}_{R} | \mb{b}_{S}\rp \\  &\propto \mbf{1}\lp \mb{b}_{S} \in \mcal{C}_{S}^{b}\rp p\lp \mb{b}_{R} | \mbf{b}_{S}\rp\\
&\overset{\aaaa}{=} \mbf{1}\lp \mb{b}_{S} \in \mcal{C}_{S}^{b}\rp \mbf{1}\lp \mb{b}_{R} \in \mcal{C}_{R}^{b}\rp p\lp \mb{b}_{R} | \mb{b}_{S}\rp.
\end{align*}
(a) is due to the fact that $\mbf{b}_{R}$ must be a codeword in $\mcal{C}_{R}^{b}$.

\begin{figure}[htbp] 
{\centering
\includegraphics[width=3.5in]{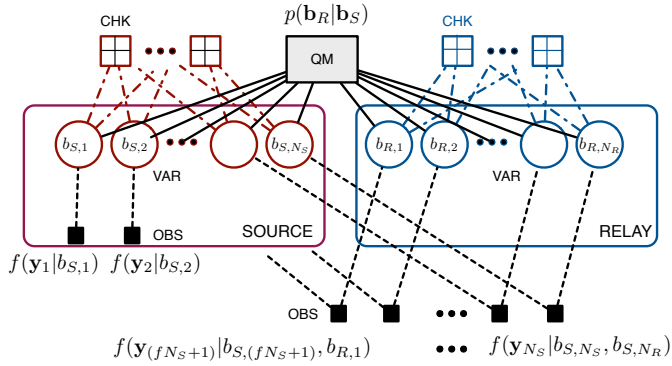}
\caption{Factor Graph for Joint Decoding.}
\label{fig_graph1}
}
\end{figure}

The resulting factor graph in Fig.~\ref{fig_graph1} shows that in addition to nodes representing channel observations i.e. $f(\mbf{y}_{i}|b_{S,i},b_{R,j})$ the subgraphs $ \mbf{1}\lp \mb{b}_{S} \in \mcal{C}_{S}^{b}\rp$ and  $\mbf{1}\lp \mb{b}_{R} \in \mcal{C}_{R}^{b}\rp$ are connected by $p\lp \mb{b}_{R} | \mb{b}_{S}\rp$ that represents the quantization operation at $R$.

If the component codes $\mathcal{C}_{S}^{b}$ and $\mathcal{C}_{R}^{b}$ are sparse, the overall factor graph is also sparse. A sum-product algorithm for decoding over such a factor graph has complexity that grows linearly with the length of component codes.  
However, the sum-product update rules at the function node $p\lp \mb{b}_{R} | \mb{b}_{S}\rp$ is very complex due to its high degree ($N_{S}+N_{R}$). Moreover, it introduces very short cycles in the graph, which deteriorates the performance of sum-product decoding. In order to get reasonably close to MAP performance and low decoding complexity, the $p\lp \mb{b}_{R} | \mb{b}_{S}\rp$ node must be factorized further. In the following subsections, choice for component codes $\mathcal{C}_{S}^{b}$ and $\mathcal{C}_{R}^{b}$ and specific techniques for factorization are discussed.


\subsection{Choice of component codes}
In the discussion above, general binary linear codes $\mcal{C}_{S}^{b}$ and $\mcal{C}_{R}^{b}$ are considered. A natural choice is to use sparse graph codes (like LDPC) that are known to have good performance and linear complexity decoding/encoding operations. 

In a previous communication \cite{5706940}, preliminary results for such factor graphs were presented using off-the-shelf LDPC codes at both $S$ and $R$.  As observed in \cite{5706940}, off-the-shelf (point-to-point) LDPC codes do not allow close-to-optimal performance over cooperative channels. An information-theoretic understanding of this observation is presented in \cite{DBLP:journals/corr/abs-1008-1766}. The authors point out that capacity-achieving codes for the point-to-point channel exhibit higher estimation errors whenever the SNR is below the Shannon limit. Therefore, in cooperative networks where the operating SNR is below the point-to-point Shannon limit, such off-the-shelf codes are no longer suitable to utilize side information from the relay at the destination. As a consequence, specialized codes are required for cooperative channels. 
For sparse graph codes, specialized code profiles that are optimized for relaying can be designed using standard tools such as density evolution analysis \cite{910578,910577}. 
However, for the LDPC-LDPC combination \cite{5706940} density evolution does not extend readily to QMF joint factor graphs. 

In this paper, the use of LDPC codes at $S$ and LDGM codes at $R$ is proposed. LDPC codes are known to perform very close to information theoretic limits when used for channel coding. Similarly LDGM codes are commonly used for lossy data compression \cite{DBLP:journals/corr/cs-IT-0408008} and the LDPC-LDGM combination is a good fit for the QMF relay channel. Moreover, density evolution analysis tools can be extended to LDPC-LDGM joint factor graphs. Such an extension is developed in Sec.~\ref{sec_densityevolution}. This permits explicit construction of code profiles optimized for relaying.  

Based on the LDPC-LDGM choice, let us introduce auxiliary variable nodes $\mbf{b}_{Q}=\{b_{Q,i}\}_{i=1}^{K_{R}}$ in the factor graph. $\mbf{b}_{Q}$ represents the $K_{R}$ bits after quantization at $R$. These are mapped to the codeword $\mbf{b_{R}}$ of length $N_{R}$ obtained after passing through a low density generator matrix having $K_{R}$ rows, $N_{R}$ columns and characteristic function $\mbf{1}(\mbf{b}_{R}\in\mcal{C}_{R}^{b}$). Since $\mbf{b}_{R}$ is a deterministic function of $\mbf{b}_{Q}$, $p\lp \mb{b}_{R} | \mb{b}_{S}\rp$ can be factorized as follows (Fig~\ref{fig_FactorGraph}):
\begin{align*}
p(\mbf{b}_{R}|\mbf{b}_{S})&=p(\mbf{b}_{R},\mbf{b}_{Q}|\mbf{b}_{S}) \\
&=p(\mbf{b}_{R}|\mbf{b}_{Q},\mbf{b}_{S})p(\mbf{b}_{Q}|\mbf{b}_{S})\\
&=\mbf{1}(\mb{b}_{R} \in \mcal{C}^{b}_{R})p(\mbf{b}_{Q}|\mbf{b}_{S})
\end{align*}
The LDGM mapping can either \emph{compress} or \emph{expand} the $K_{R}$ quantized bits i.e. the LDGM coding rate can be greater than $1$. The $\mbf{b}_{R}$ nodes always have degree $2$ and they simply perform forwarding of messages under the sum-product algorithm.

\subsection{Scalar Quantizer}
In general, a vector quantizer can be used at $R$. However, it is shown \cite{2009arXiv0906.5394A} that QMF performs within bounded gap of capacity even with a scalar quantizer. Under scalar quantization, the observation for every bit from $S$ is quantized independently. If each $y_{R,i}$ is quantized into $b_{Q}[A_i]$ for $i=1,2,\ldots fN_{S}$, then the $p\lp \mb{b}_{Q} | \mb{b}_{S}\rp$ function node factorizes into $fN_{S}$ separate nodes each representing a scalar quantization operation. 
\begin{align*}
p\lp \mb{b}_{Q} | \mb{b}_{S}\rp &= \prod_{i=1}^{f N_{S}} p\lp b_{Q}[A_i]| b_{S,i}\rp,\ \text{where} \\
\bigcup_{i=1}^{fN_{S}}A_i &= \{1,2,\ldots, K_R\},\ A_i \cap A_j = \emptyset\ \forall i\ne j 
\end{align*}
where $A_i$ denotes the subset of indices in $\mb{b}_{Q}$ that observation $y_{R,i}$ is quantized into.
\begin{figure}[htbp] 
{\centering
\includegraphics[width=3.5in]{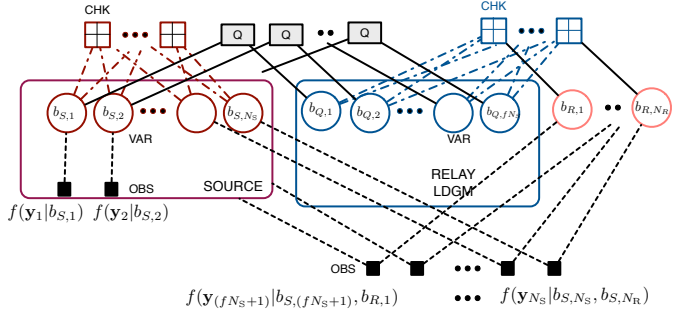}
\caption{Factor graph: LDPC code at $S$, LDGM code at $R$ with $1$ bit scalar quantizer.}
\label{fig_FactorGraph}
}
\end{figure}
As a result, the variable nodes of the two Tanner graphs are connected by function nodes representing the stochastic relations $p\lp b_{Q}[A_i]| b_{S,i}\rp$ among them. Henceforth, these are called \emph{quantize} (Q) nodes, as they are induced by the quantization procedure at the relay. An example factor graph showing the LDPC-LDGM construction is illustrated in Fig~\ref{fig_FactorGraph} where each symbol observation $y_{R,i}$ is quantized into one bit (i.e. $K_{R}=fN_{S}$ and $A[i]=\{i\}$).
As shown, there are four kinds of nodes in the resulting factor graph: observation (OBS) nodes, variable (VAR) nodes, check (CHK) nodes, and quantize (Q) nodes. 
Some VAR nodes in the $\mathcal{C}_{S}^{b}$ subgraph share OBS nodes with VAR nodes in the $\mathcal{C}_{R}^{b}$ subgraph. This is because of multiple access at $D$.

\subsection{DBLAST Scheme}
\label{subsec_dblast}
\begin{figure}[htbp]
{\center
\includegraphics[width=3in]{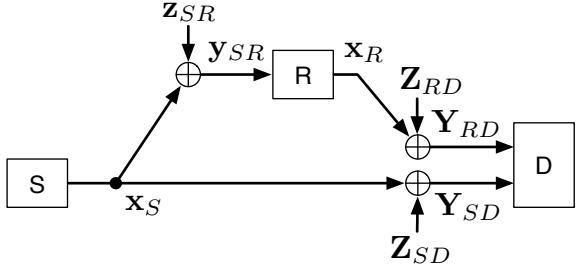}
\caption{Channel model using DBLAST}
\label{fig_EquivCh}
}
\end{figure}
The factor graph shown in Fig.~\ref{fig_FactorGraph} can be simplified further using the Diagonal Bell Labs Space-Time architecture (DBLAST) \cite{BLTJ:BLTJ2015}.
As discussed in the Appendix, the degree $2$ OBS nodes (representing multiple access) are factorized using DBLAST. 
Under DBLAST, the destination observes two orthogonal sets of observations (see Fig.~\ref{fig_EquivCh}). The factorization is shown in equation \eqref{long_eq} where ~$\ul{\mb{Y}}~:=~[\mb{Y}_{SD}\ \mb{Y}_{RD}]$.
\begin{figure*}
\normalsize  
\begin{align}
\label{long_eq}
f(\ul{\mb{Y}}| \mb{b}_{S}, \mb{b}_{Q}) &= \prod_{i=1}^{fN_{S}} f(\mbf{y}_{SD,i}|b_{S,i}) \prod_{j=1}^{N_{R}} f(\mbf{y}_{SD, (f N_{S}+j)}, \mbf{y}_{RD, (f N_{S}+j)}|b_{Q,j},b_{{S},(f N_{S}+j)}) \notag \\
&= \prod_{i=1}^{fN_{S}} f(\mbf{y}_{SD,i}|b_{S,i}) \prod_{j=1}^{N_{R}} f(\mbf{y}_{SD, (f N_{S}+j)}|b_{{S},(f N_{S}+j)})f(\mbf{y}_{RD, (f N_{S}+j)}|b_{Q,j}) \notag \\
&= \prod_{i=1}^{N_{S}} f(\mbf{y}_{SD,i}|b_{S,i}) \prod_{j=1}^{N_{R}} f(\mbf{y}_{RD, (f N_{S}+j)}|b_{Q,j}),
\end{align}
\hrulefill
\vspace*{4pt} 
\end{figure*}

\begin{figure}[htbp]
{\center
\includegraphics[width=3.5in]{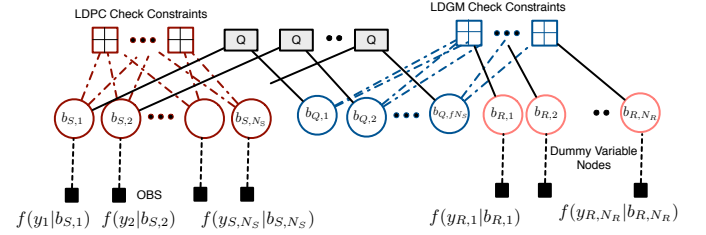}
\caption{Simplified factor graph with one-bit scalar quantizer and DBLAST.}
\label{fig_FactorGraph2}
}
\end{figure}

An example of the simplified factor graph is depicted in Fig.~\ref{fig_FactorGraph2}. In this graph, VAR nodes in the two Tanner graphs are connected \emph{only} through Q nodes. Since $\mbf{y}_{RD,i}=\mbf{0}$ for $i=1,\ldots, fN_{S}$, we rename $\mbf{y}_{RD, (f N_{S}+j)} \equiv \mbf{y}_{RD, j}$, for $j=1,\ldots,N_{R}$.

The resulting graph has a structure similar to an irregular LDPC code but with special Q constraints. In Sec~\ref{sec_decoding}, sum-product updates for this graph are derived following the general principle outlined in \cite{910572}. It is shown that for a simple one-bit quantizer each Q node further factorizes into a CHK constraint and a dummy VAR node. This reduces the factor graph to a Tanner graph that does not have any special nodes. Such a property is useful to leverage existing techniques used for the design of low-power, high-throughput LDPC decoders.




\subsection{Decoding Algorithm}
\label{sec_decoding}
For the point-to-point system, \emph{belief-propagation} is an iterative algorithm that computes the {\it a posteriori} probability to decode message bits. The algorithm computes this exactly if the factor graph has no cycles. Otherwise, it computes the approximate {\it a posteriori} probability for each bit \cite{910572}. For the factor graph in Fig~\ref{fig_FactorGraph2}, messages being passed on the edges of the factor graph and the update rules at the variable/check nodes stay unchanged. The only new ingredient in the mix is the Q nodes introduced by our framework. 

Let the subscripts $\msf{V}$, $\msf{C}$, and $\msf{Q}$ denote VAR nodes, CHK nodes, and Q nodes respectively. For $\msf{F}\in \{\msf{C}, \msf{Q}\}$, let $\omega_{\msf{V}\msf{F}}^{(l)}$ denote the message sent from variable node $\msf{V}$ to function node $\msf{F}$ in the $l^{\text{th}}$ iteration. Every edge in the graph is connected to exactly one variable node and a message on the edge represents the {\it a posteriori} probability for the respective variable. The messages can be represented as LLRs, but for the sake of simplicity we represent the messages as a two-dimensional vector in this subsection\footnote{Later we will replace $\omega$ by $w$, the commonly used message $\log\frac{p_{0}}{p_{1}}$ (LLR) in belief propagation.}.
$\omega_{\msf{V}\msf{F}}:=[p_0 \ p_{1}] $,
where $\omega_{\msf{V}\msf{F}}(1)=p_{0} \in [0,1]$ represents the probability that the bit is $0$ and $\omega_{\msf{V}\msf{F}}(2)=p_{1} \in [0,1]$ represents the probability that the bit is $1$ $(p_{0}+p_{1}=1)$.


The message sent from $\msf{V}$ to $\msf{F}\in \{\msf{C}, \msf{Q}\}$ is the normalized product of all incoming messages into $\msf{V}$ except for the message from $\msf{F}$. The normalization ensures that $p_{0}+p_{1}=1$ for the outgoing message. 
The message sent from $\msf{C}$ to $\msf{V}'$ is the indicator function that the check is satisfied, marginalized on the bit represented by $\msf{V}'$. 
The message sent from $\msf{Q}$ to $\msf{V}$ is the marginalization of the function $p\lp b_{Q}[A_i] |b_{S,i}\rp$ on the symbol represented by $\msf{V}$. $b_{Q}$ is computed from a noisy observation of $b_{S,i}$, the node $\msf{Q}$ imposes a probabilistic constraint on the variables. Since the quantization is scalar: $\forall \mb{u}\in\{0, 1\}^{|A_i|}$ and $v\in\{0, 1\}$,
\begin{align*}
g(\mb{u},v) := p\lp b_{Q}[A_i]= \mb{u} |b_{S,i}=v\rp.
\end{align*}
This function is fully represented by a lookup table with $2^{|A_i|+1}$ values, which is used to derive the update rule for $\msf{Q}$.

\begin{figure}[!t]
{\centering
\includegraphics[width=3.5in]{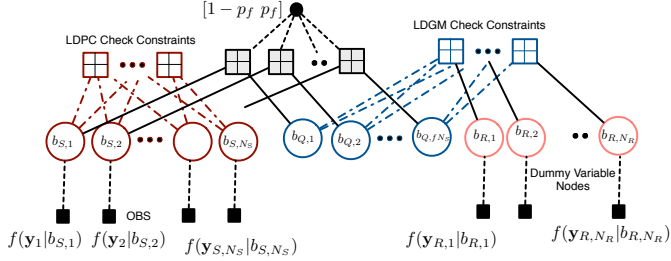}
\caption{Equivalent factor graph of that in Fig. \ref{fig_FactorGraph2}. The $\msf{Q}$ nodes are factorized into a CHK node and a dummy variable.}
\label{fig_FactorGraph3}
}
\end{figure}

As an example, let us consider a one-bit scalar quantizer at the relay and derive the update rule. For this case, the Q node can be further factorized into a CHK node  and a dummy VAR node that sends a constant message. Note that 
$
A_i = \{i\},\ i=1,2,\ldots, fN_{S}
$
and $K_{R} = fN_{S}$. The factor graph is depicted in Figure \ref{fig_FactorGraph2}.  Consider $\forall u\in\{0, 1\}$ and $v\in\{0, 1\}$,
\begin{align*}
g(u,v) :=& p\lp b_{Q,i} = u |b_{S,i}=v\rp \notag \\ =& (1-p_f)\mbf{1}\{u=v\} + p_f\mbf{1}\{u\ne v\}
\end{align*}
here $p_f := \frac{1}{2}\erfc{\sqrt{\frac{\SNR_{SR}}{2}}}$ denotes the probability of bit error for scalar one-bit quantization over a BIAWGN channel. Since the function $g$ is symmetric in $u$ and $v$, it can be assumed that the VAR node is of the source, and the marginalization is on $v$. Let the other VAR node be $V'$. This leads to the following update rule:
\begin{align*}
\omega_{\msf{QV}}(1) &= (1-p_f)\omega_{\msf{V}'\msf{Q}}(1)+p_f\omega_{\msf{V}'\msf{Q}}(2)\\
\omega_{\msf{QV}}(2) &= (1-p_f)\omega_{\msf{V}'\msf{Q}}(2)+p_f\omega_{\msf{V}'\msf{Q}}(1),
\end{align*}
This takes the same form of a CHK node update with incoming messages $\omega_{\msf{V}'\msf{Q}}$ and $[1-p_f\ p_f]$. Therefore, the Q node in this set-up specializes to a CHK node with additional dummy VAR nodes sending constant message $[1-p_f\ p_f]$ that depends on $\SNR_{SR}$. The resulting factor graph is depicted in Fig. \ref{fig_FactorGraph3}.

\section{Code Design}
\label{sec_densityevolution}
In this section, design of specific codes for QMF relaying is discussed. Typically sparse graph codes like LDPC and LDGM are drawn randomly from ensembles, which are described using degree profiles. In the point-to-point case, if the block-length is sufficiently large, the decoding performance of such codes converges to the \emph{ensemble average} \cite{910577}. Let us consider degree profiles $(\lambda_{S},\rho_{S})$ and $(\lambda_{R},\rho_{R})$ for the LDPC and LDGM codes at source and relay respectively. $\lambda_{S}$ and $\rho_{S}$ are polynomials representing variable and check degree distributions for the LDPC code:
\begin{align*}
\lambda_{S}(x) = \sum_{i=2}^{\infty} \lambda_{S,i}x^{i-1},\ \rho_{S}(x) = \sum_{i=2}^{\infty} \rho_{S,i}x^{i-1}
\end{align*}
Here $\lambda_{S,i}$ and $\rho_{S,i}$ denote the fraction of edges with degree $i$ at a variable node and at a check node respectively. For the LDGM code, we have the similar definition for $\lambda_{R}$ and $\rho_{R}$ except that these are regarding the edges connecting check nodes and variable nodes for $\mb{b}_{Q}$ (not $\mb{b}_{R}$).

These profiles must satisfy the following constraints:
\begin{align*}
&\mathcal{R} =1-\frac{\int_{0}^{1}\rho_{S}(x)dx}{\int_{0}^{1}\lambda_{S}(x)dx},\ 
\frac{K_{R}}{N_{R}} =\frac{\int_{0}^{1}\rho_{R}(x)dx}{\int_{0}^{1}\lambda_{R}(x)dx} =\frac{f}{1-f},\\
&\lambda_S(1) = \lambda_R(1) = \rho_S(1) = \rho_R(1) = 1
\end{align*}
In addition to the sub-graphs representing the two component codes, the joint factor graph shown in Fig.~\ref{fig_FactorGraph3} also includes edges connecting them (via Q nodes). As discussed previously in Sec~\ref{sec_coding}, let us consider a fixed one bit scalar quantizer. The edges connecting with Q nodes are considered \emph{fixed} in the rest of this section. In contrast, the edges in LDPC and LDGM subgraphs are drawn randomly from the ensemble using construction procedure described in \cite{910577}.

In point-to-point channels, the typical method to analyze and design sparse graph codes is to compute the ensemble average performance for given degree profiles assuming infinite block length (\emph{convergence to computation trees}). The ensemble average performance (decoding error probability for given SNRs) is calculated using \emph{density evolution} developed in \cite{910578,910577}. The two key elements of classical density evolution, namely, \emph{concentration around ensemble average} and \emph{convergence to computation tree channels} for sufficiently large block length hold for the proposed QMF relaying system as well. The proofs can be readily extended from those of point-to-point channels \cite{910577} \cite{910578}. 

Without loss of generality it is assumed that the all-zero codeword is transmitted from $S$. This is a result of the symmetry of the relay channel. 
However, $R$ does not transmit the all-zero codeword because the source-to-relay channel is noisy. For sufficiently large block lengths and a given value of $\SNR_{SR}$ there is a \emph{typical} sequence $\mbf{b}_{Q}$ that is mapped to a \emph{typical} $\mbf{b}_{R}$ based on the LDGM code. The probability of occurrence for atypical codewords vanishes as the block length becomes large. Therefore, it is ignored for computing the ensemble average performance. A typical $\mbf{b}_{Q}$ comprises of $K_R(1-p_f)$ $0$'s and $K_Rp_f$ $1$'s. $p_{f}$ is defined in Sec.~\ref{sec_decoding}. For a given degree profile, $\lp\lambda_{R}, \rho_{R}\rp$, each bit of the typical $\mbf{b}_{R}$ is i.i.d. Bernoulli$(q)$, where $q$ is the probability of having an odd number of $1$'s in a column of the generator matrix (drawn randomly from the LDGM ensemble).
\begin{align*}
q = \sum_j \lp \frac{\rho_R(j)/j}{\sum_i\rho_R(i)/i}\rp\frac{1-(1-2p_f)^j}{2}
\end{align*}




To develop density evolution rules for QMF relaying, we consider the belief-propagation algorithm with log-likelihood ratios as the messages passed among variable nodes and various function nodes.
Let $w_{\msf{F}\msf{V}}^{(l)}$ and $w_{\msf{V}\msf{F}}^{(l)}$ denote the message sent from the function node $\msf{F}$ to the variable node $\msf{V}$ and vice-versa, at the $l$-th iteration. $\msf{F}\in\{ \msf{C}_S, \msf{C}_R, \msf{Q},\msf{O}_S, \msf{O}_R\}$ represent the LDPC CHK nodes, LDGM CHK nodes, Q nodes, OBS nodes at $S$ and $R$ respectively. $\msf{V}\in\{ \msf{V}_S, \msf{V}_Q, \msf{V}_R\}$ represent the VAR nodes corresponding to $\mbf{b}_{S},\mbf{b}_{Q}$ and $\mbf{b}_{R}$ respectively.

%
The sum-product update rules in terms of the commonly used LLR's are written as follows: \begin{align}
w_{\msf{V}\msf{F}}^{(l)} &= \sum_{\msf{F}'\in \mcal{N}(\msf{V})\setminus\{\msf{F}\}} w_{\msf{F}'\msf{V}}^{(l)} \label{eq_vtof}\\
w_{\msf{O}\msf{V}}^{(l)} &= w_{\msf{V}}, \quad (\msf{O},\msf{V}) = \{(\msf{O}_S,\msf{V}_S), (\msf{O}_R,\msf{V}_R)\} \notag \\
w_{\msf{F}\msf{V}}^{(l+1)} &= 2\tanh^{-1}\lp \prod_{\msf{V}'\in\mcal{N}(\msf{F})} \tanh\lp \frac{1}{2}w_{\msf{V}'\msf{F}}^{(l)}\rp\rp \label{eq_ftov}\\ 
&\quad \text{if } \msf{F} = \msf{C}_S,\msf{C}_R \notag\\
w_{\msf{F}\msf{V}}^{(l+1)} &= 2\tanh^{-1}\lp (1-2p_f)\prod_{\msf{V}'\in\mcal{N}(\msf{F})} \tanh\lp \frac{1}{2}w_{\msf{V}'\msf{F}}^{(l)}\rp\rp \notag\\ 
&\quad \text{if } \msf{F} = \msf{Q} \notag
\end{align}
Here $\mcal{N}(\cdot)$ here denote the set of neighboring nodes and $w_{\msf{V}}$ represents the LLR from channel observation.

Compared to the point-to-point case where there is only one kind of variable node ($\msf{V}$) and one kind of CHK node ($\msf{C}$) the update rules can be expressed simply by \eqref{eq_vtof} and \eqref{eq_ftov} where $\msf{F}=\msf{C}$. Density evolution analysis tracking the density of these messages in each iteration. For the point-to-point case with degree distribution $(\lambda,\rho)$, there is only one type of edge and the evolution is expressed using a pair of coupled recursive equations as follows: 
\begin{align*}
P_{\msf{C}\msf{V}}^{(l+1)} &= \Gamma^{-1}\lp \sum_{j} \rho_j \lp\Gamma\lp P_{\msf{V}\msf{C}}^{(l)}\rp\rp^{\otimes(j-1)} \rp\\
P_{\msf{V}\msf{C}}^{(l)} &= P_{\msf{V}}\otimes\sum_{i}\lambda_i \lp P_{\msf{C}\msf{V}}^{(l)}\rp^{\otimes(i-1)}
\end{align*}
Here  $\Gamma\lp\cdot\rp$ denotes a transformation on the density as defined in \cite{910578}, $\otimes$ denotes the convolution operator and $P^{(l)}_{\{\cdot\}}$ denotes the density of message $w^{(l)}_{\{\cdot\}}$. $P_{\msf{V}}$ represents the conditional density of the LLR of the point-to-point channel.

For the QMF relaying case, there are $4$ types of edges and densities for messages along all of them must be tracked. The recursive density updates are derived similarly:

\subsubsection*{Function nodes to variable nodes}
\begin{align*}
&P_{\msf{C}_S\msf{V}_S}^{(l+1)} = \Gamma^{-1}\lp \sum_{j} \rho_{S,j} \lp\Gamma\lp P_{\msf{V}_S\msf{C}_S}^{(l)}\rp\rp^{\otimes(j-1)} \rp\\
&P_{\msf{C}_R\msf{V}_Q}^{(l+1)} = \Gamma^{-1}\lp \sum_{j} \rho_{R,j} \lp\Gamma\lp P_{\msf{V}_Q\msf{C}_R}^{(l)}\rp\rp^{\otimes(j-1)}\otimes \Gamma\lp P_{\msf{V}_R\msf{C}_R}^{(l)}\rp \rp\\
&P_{\msf{Q}\msf{V}_S}^{(l+1)} = \Gamma^{-1}\lp \Gamma\lp P_{\msf{V}_Q\msf{Q}}^{(l)}\rp\otimes\Gamma\lp \delta_{\log\frac{1-p_f}{p_f}}\rp \rp\\
&P_{\msf{Q}\msf{V}_Q}^{(l+1)} = \Gamma^{-1}\lp \Gamma\lp P_{\msf{V}_S\msf{Q}}^{(l)}\rp\otimes\Gamma\lp \delta_{\log\frac{1-p_f}{p_f}}\rp \rp
\end{align*}
\subsubsection*{Variable nodes to function nodes}
\begin{align*}
P_{\msf{V}_S\msf{C}_S}^{(l)} =& P_{\msf{V}_S}\otimes\sum_{i}\lbp f\lambda_{S,i} \lp P_{\msf{C}_S\msf{V}_S}^{(l)}\rp^{\otimes(i-1)}\otimes P_{\msf{Q}\msf{V}_S}^{(l)} + \right. \\ & \left.(1-f)\lambda_{S,i} \lp P_{\msf{C}_S\msf{V}_S}^{(l)}\rp^{\otimes(i-1)}\rbp\\
P_{\msf{V}_Q\msf{C}_R}^{(l)} =& \sum_{i}\lambda_{R,i} \lp P_{\msf{C}_R\msf{V}_Q}^{(l)}\rp^{\otimes(i-1)}\otimes P_{\msf{Q}\msf{V}_Q}^{(l)}\\
P_{\msf{V}_R\msf{C}_R}^{(l)} =& P_{\msf{V}_R}\\
P_{\msf{V}_S\msf{Q}}^{(l)} =& P_{\msf{V}_S}\otimes\sum_{i}\lambda_{S,i} \lp P_{\msf{C}_S\msf{V}_S}^{(l)}\rp^{\otimes(i)}\\
P_{\msf{V}_Q\msf{Q}}^{(l)} =& \sum_{i}\lambda_{R,i} \lp P_{\msf{C}_R\msf{V}_Q}^{(l)}\rp^{\otimes(i)}
\end{align*}

Here $\delta_r(\cdot)$ denotes the Dirac delta function at point $r\in\mbb{R}$. $\delta_{\log\frac{1-p_f}{p_f}}$ shows up in the expressions because the Q node is equivalent to a CHK node connected to a \emph{constant}. The differences in evolution rules between the QMF relaying and point-to-point channel arise due to the probabilistic Q constraints in the joint factor graph. 





As in the point-to-point case, $P_{\msf{V}_S}$ is the conditional density of the LLR of the  source to destination channel, given that an all-zero codeword is sent from $S$. $P_{\msf{V}_R}$ is the marginal density of the LLR of the relay to destination channel under the marginal law that $\mbf{b}_{R}$ is i.i.d. Bernoulli$(q)$.

The density evolution rules derived above are used to compute the probability of error in decoding of $\mbf{b}_{S}$. For successive interference cancellation using DBLAST, $\mbf{b}_{R}$ must also be reliably decoded. Density evolution rules to compute probability of decoding error for $\mbf{b}_{R}$ can be similarly derived.

\section{Bit Interleaved Coded Modulation} \label{subsec_BICM}
So far, the discussion has focussed on the BMS Gaussian relay defined in Sec.~\ref{sec_problem}. For the high SNR regime, input alphabet $\mbf{x}_{S}\in{\mathcal{A}^{N_{S}}}$ and $\mathbf{x}_{R}\in\mathcal{A}^{N_{R}}$ where $\mathcal{A}$ represents constellation points in a high-order modulation scheme must be considered. In practice many systems use BICM \cite{669123} to combine channel codes designed for binary alphabet with high-order signal constellations. BICM has also been proposed for various cooperative channel scenarios \cite{1600070}\cite{4313147}\cite{4939335}\cite{4259765}. 
In this subsection, a procedure is discussed for extending the coding framework from the BMS relay channel to a relay channel with inputs from high-order alphabets.

Under classical BICM \cite{669123}, a point to point Gaussian channel is decomposed into \emph{parallel independent memoryless} ``sub-channels''. Every ``sub-channel'' $p_{Y|B,{S}}(y|b,s)$ has binary inputs $b\in\{0,1\}$ and depends on state $s\in\{1,2,\ldots,L\}$ which is chosen uniformly and known to both the terminals ($2^{L}$ is the cardinality of the chosen signal constellation). At the receiver, LLR for a bit that was mapped to state $s$ is calculated from symbol observation $y\in\mathbb{C}$ (in case of MIMO receiver $\mbf{y}\in\mathbb{C}^{M}$).
\begin{align*}
LLR(y,s)=\log\frac{P_{B|Y,S}(b=0|y,s)}{P_{B|Y,S}(b=1|y,s)}
\end{align*}

\begin{figure}[htbp]
{\centering
\subfigure[PBICM Transmitter]{\includegraphics[width=3.5in]{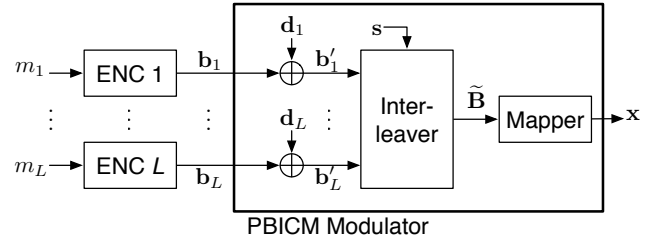}}
\subfigure[PBICM Receiver]{\includegraphics[width=3.5in]{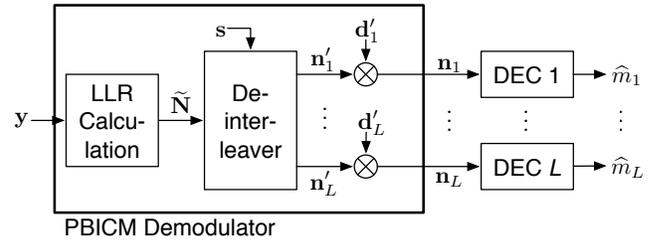}}
\caption{PBICM architecture. $\{\mb{d}_i\}_{i=1}^L$ are dithers, and $\mb{d}_i' = 1-2\mb{d}_i$.} 
\label{fig_BICM}
}
\end{figure}

However, this binary channel is not guaranteed to be output-symmetric i.e. the crossover probability for a bit is not independent of its value. Let $f_{\Lambda}(\lambda)$ represent the PDF of $LLR(y,s)$. The channel is output symmetric if the following condition holds:
\begin{align*}
f_{\Lambda|B}(\lambda|b=0)=f_{\Lambda|B}(-\lambda|b=1)
\end{align*}
Conventional methods for designing linear coding schemes such as density evolution etc. cannot be used with asymmetric channels. This  issue is resolved by adding random dithers at every bit to make the channel output-symmetric as proposed in \cite{669123,HouSiegel_03,IngberFeder_10}. Dithers are i.i.d. $\mathrm{Bernoulli}\lp\frac{1}{2}\rp$ variables known to both the transmitter and receiver. For a dither $d\in\{0,1\}$ the channel $p_{Y|B,{S},{D}}(y|b,s,d)$ is binary, memoryless and symmetric (BMS). 
\begin{align*}
LLR(y,s,d)=(-1)^{d}LLR(y,s)
\end{align*}
This method is called parallel BICM (PBICM) in \cite{IngberFeder_10} and Fig.~\ref{fig_BICM} shows the architecture for a PBICM point to point link having $L$ states i.e. signal constellation of size $2^{L}$. $\{m_{i}\}_{i=1}^{L}$ represent messages and $\mbf{b}_i$ and $\mbf{b'}_{i}$ the transmit codewords before and after dithering. The equivalent BMS channel can be characterized by $L$, the SNR of the underlying AWGN channel and the symbol mapping in modulation. In the rest of this paper we consider that  \emph{Gray} mapping is used.

In order to use PBICM with the relay channel, a definition of quantize-and-map operation under PBICM is required. With a PBICM modulator at source $S$, the observations at relay $R$ ($\mbf{y}_{R}$) represent $L$ interleaved codewords. If $R$ performs quantization at the \emph{symbol} level, then the decomposition into independent binary sub-channels will be lost. As an alternative, it is proposed that $R$ perform quantization at the \emph{bit} level. 

\begin{figure}[htbp]
{\centering
\includegraphics[width=3.5in]{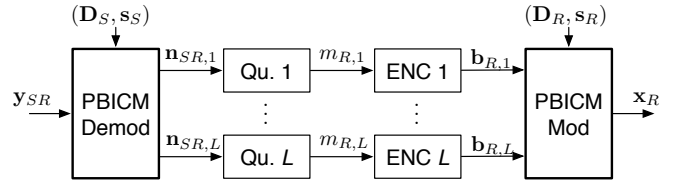}
\caption{QMF relaying with PBICM.}
\label{fig_BICM_QMF}
}
\end{figure}

\begin{figure*}[!t]
{\centering
\subfigure[Equivalent relay channel. \label{fig_BICM_Relay_sub}]{\includegraphics[width=3in]{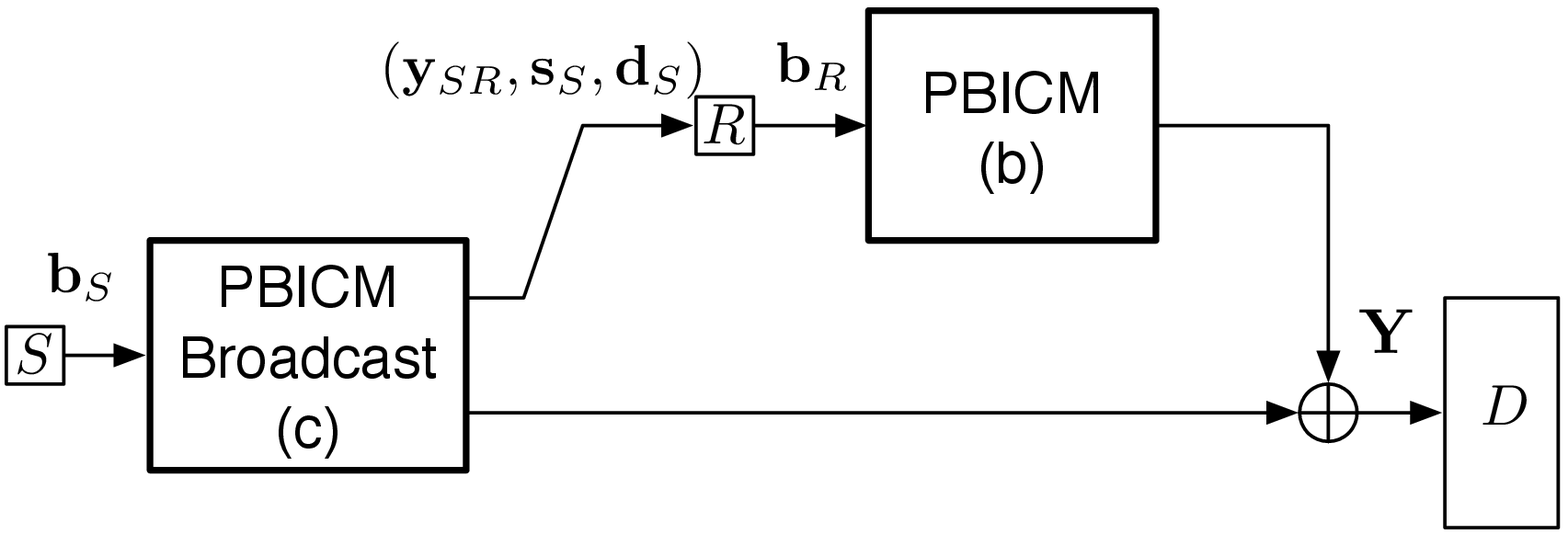}}
\subfigure[Equivalent point-to-point channel.]{\includegraphics[width=2in]{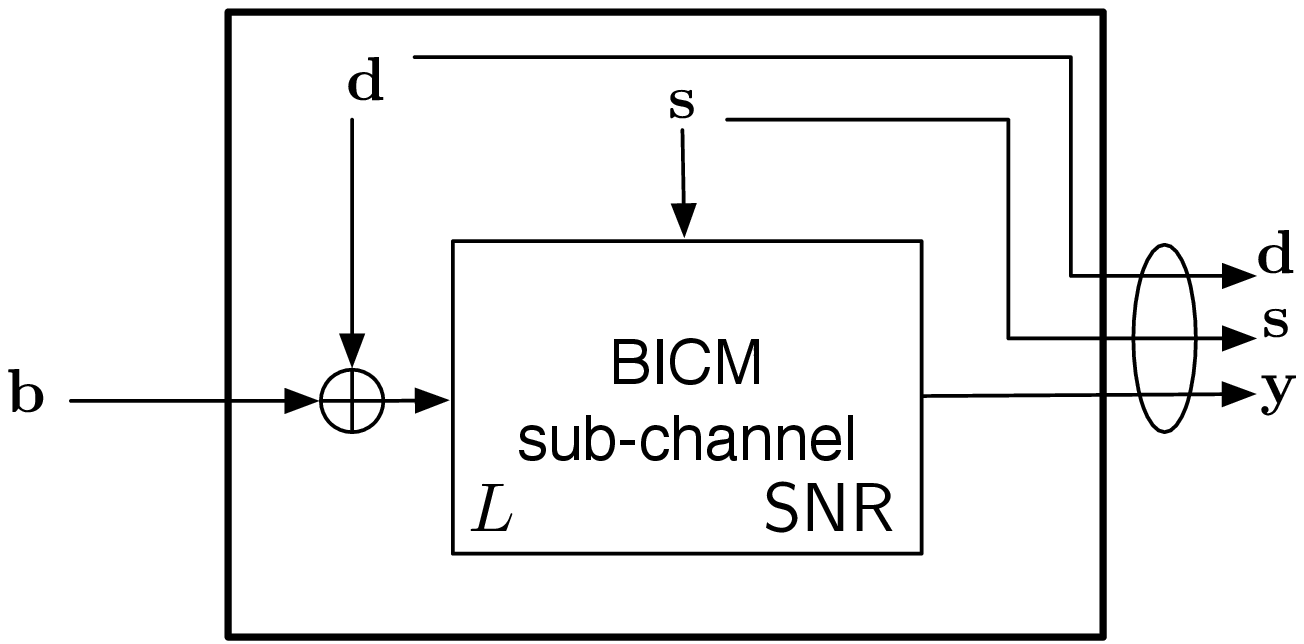}}
\subfigure[Equivalent broadcast channel.]{\includegraphics[width=2in]{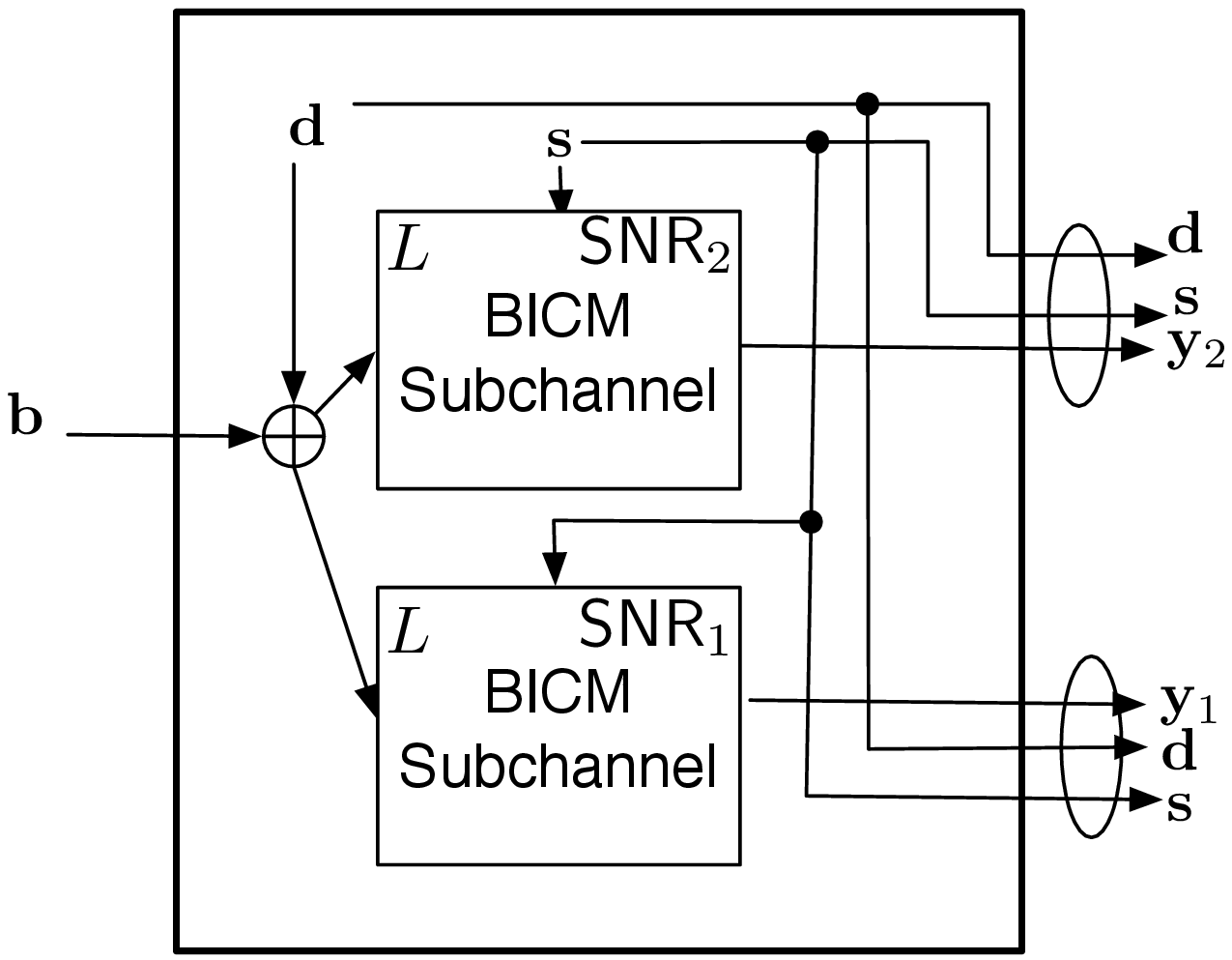}}
\caption{Equivalent binary-input system.}
\label{fig_BICM_Relay}
}
\hrulefill
\end{figure*}

$S$ and $R$ both use PBICM modulator blocks with constellation size $2^{L}$ having state and dither vectors given by $\mbf{s}_{S},\mbf{s}_{R},\mbf{D}_{S}$ and $\mbf{D}_{R}$ respectively. The QMF operation at $R$ is described below (depicted in Fig.~\ref{fig_BICM_QMF}):
\begin{IEEEenumerate}
\item  For observed symbol sequence $\mb{y}_{SR} := \{y_{SR,j}\}_{j=1}^{fN_{S}}$ perform PBICM demodulation. The output is represented as $\{\mbf{n}_{SR,i}\}_{i=1}^{L}$ where each $\mb{n}_{SR,i} :=\{n_{SR,i,j}\}_{j=1}^{fN_{S}}$ represents LLRs for the $i^{\text{th}}$ codeword.
\item  Quantize every LLR in $\{\mbf{n}_{SR,i}\}_{i=1}^{L}$. As an example, for a one bit scalar quantizer this simply involves observing the sign of LLRs.
\item  Encode the quantizer output $\{m_{R,i}\}_{i=1}^{L}$ using an LDGM code.
\item  Transmit the resultant codewords $\{\mbf{b}_{R,1}\}_{i=1}^{L}$ using a PBICM modulator.
\end{IEEEenumerate}
Using this definition of QMF, the Gaussian relay channel is decomposed into parallel BMS relay channels.
The BMS relay channel is shown in Fig.~\ref{fig_BICM_Relay}. It is characterized by constellation size at $S$ and $R$ and the $\SNR$ of the underlying AWGN links i.e. $\SNR_{SR},\SNR_{SD},\SNR_{RD}$.


\section{Link Design Example}
\label{sec_example}
In Sec.~\ref{sec_densityevolution} an extension of density evolution tools was developed \cite{910578}\cite{910577} for joint LDPC-LDGM factor graphs based on QMF relaying. In this section, a link design example with construction of explicit codes is shown for a DBLAST-equivalent channel shown in Fig.~\ref{fig_EquivCh} BMS relay channel. The performance of designed codes is presented using simulations with high order modulation based on PBICM principles described in Sec.~\ref{subsec_BICM}.

\subsection{System Parameters}
The capacity advantage of cooperative relaying is most pronounced when the source to relay link is significantly better than the direct link between source and destination.
We therefore consider an example scenario where the $S$ to $R$ link is $10$ dB stronger than the others.
\begin{align}
\SNR_{SD}=\SNR_{RD}, \ \ 
\SNR_{SR} =10 \times \SNR_{SD}
\label{eq_param}
\end{align} 

\subsubsection{Modulation Order}
As a guideline for system design use the following information-theoretic bound on maximal achievable rate using QMF relaying with continuous Gaussian inputs $x_{S}$ and $x_{R}$ and a vector Gaussian quantizer at the noise level. 
\begin{align}\label{eq_QMF_Rate_G}
&\mcal{R}_{\rm{QMF,G}} =\\ \notag
& \min\lbp \begin{array}{l}(1-f)\C_{\rm{G}}\lp \SNR_{SD}\rp + f\C_{\rm{G}}\lp\frac{\SNR_{SR}}{2} + \SNR_{SD}\rp,\\
(1-f)\C_{\rm{G}}\lp\SNR_{RD}\rp + \C_{\rm{G}}\lp \SNR_{SD}\rp -f
\end{array}\rbp
\end{align}
Here $\C_{\rm{G}}\lp x \rp := \log\lp 1+x \rp$ is the AWGN point-to-point capacity at signal-to-noise ratio $x$.
If the inputs are constrained to structured constellations such as $16$ QAM, $64$ QAM, then the achievable rate with $2^{2n}$-QAM modulation and BICM is computed as follows:
\begin{align}\label{eq_QMF_Rate_CM}
&\mcal{R}_{\mathrm{QMF},n} =\\ \notag
& \min\lbp \begin{array}{l}(1-f)\C_{n}\lp \SNR_{SD}\rp + f\C_{n}\lp\frac{\SNR_{SR}}{2} + \SNR_{SD}\rp,\\
(1-f)\C_{n}\lp\SNR_{RD}\rp + \C_{n}\lp \SNR_{SD}\rp -f
\end{array}\rbp
\end{align}
Here too we use a vector Gaussian quantizer at the noise level. Note that $n\in\{2,3,4\}$ and $\C_{n}\lp x \rp$ denotes the $2^{2n}$-QAM constellation-constrained point-to-point capacity at signal-to-noise ratio $x$ under BICM.
\subsubsection{Listening-time Fraction}
For QMF, the listening-time fraction $f$ at $R$ can be independently optimized to maximize system throughput \cite{4305423,5205885,4797531}. The optimal $f^*$ is found by balancing the two terms in the minimization of \eqref{eq_QMF_Rate_G}:
\begin{align*}
&(1-f^*)\C_{\rm{G}}\lp \SNR_{SD}\rp + f^*\C_{\rm{G}}\lp\frac{\SNR_{SR}}{2} + \SNR_{SD}\rp\\
&=(1-f^*)\C_{\rm{G}}\lp\SNR_{RD}\rp + \C_{\rm{G}}\lp \SNR_{SD}\rp -f^* 
\end{align*}

Alternatively a sub-optimal listening fraction $f$ can be used based on reduced channel knowledge at relay. It is shown in \cite{5205885} that this does not have a significant impact on throughput.

\begin{figure}[htbp]
{\centering
\includegraphics[width=3.5in]{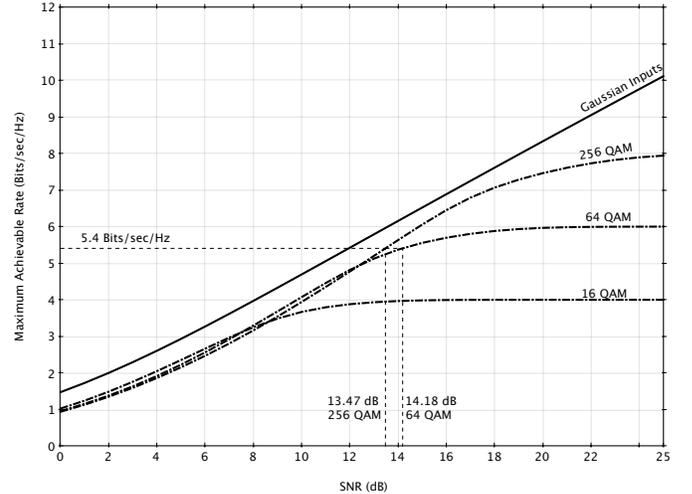}
\caption{Maximum achievable rate for QMF relaying with modulation constraints on channel inputs plotted vs $\SNR_{SD}$ for SNR relationships in Eq.~\eqref{eq_param}}
\label{fig_example_mod}
}
\end{figure}

For system parameters in Eq~\eqref{eq_param}, $\mathcal{R}_{\rm{QMF},\rm{G}}$ and $\mathcal{R}_{\mathrm{QMF},n}$ are plotted for $n=2,3,4$ in Fig~\ref{fig_example_mod} vs. $\SNR_{SD}$. For each point, the optimized listening fraction $f^*$ is used. To design a link with throughput of $5.4$ information bits per symbol, both $64$ QAM and $256$ QAM are potentially good choices for modulation having QMF information theoretic thresholds at $14.18$ dB and $13.47$ dB respectively. Let us choose $64$ QAM ($6$ coded bits per symbol) for the example design, which means that $S$ should use an LDPC code of rate $\mcal{R}=\frac{5.4}{6}=0.9$. The optimal listening fraction corresponding to $\SNR_{SD}=14.18$ dB is $f^{*}\approx\frac{2}{3}$. This determines the LDGM coding rate 
\[\frac{K_{R}}{N_{R}}=\frac{f}{1-f}\approx2\]

\subsection{Code Design}
Codes $\mcal{C}_{S}^{b}$ and $\mcal{C}_{R}^{b}$ optimized for the above system parameters can be designed using density evolution tools \cite{910578}. This involves finding good degree profiles that have the lowest possible decoding SNR threshold and randomly generating finite block length codes from them. 

In order to reduce the computational complexity of density evolution we use the Gaussian approximation to density evolution developed in \cite{910580}. Additionally, we use the following heuristics to reduce the search space for profiles. 
\begin{enumerate}
\item For $\mcal{C}_{S}^{b}$ we consider check degree profiles that are concentrated \cite{910580} i.e. all check degrees (from edge perspective) are either $k$ or $k+1$ for some integer $k\geq2$.
\item For $\mcal{C}_{S}^{b}$ we consider variable degree profiles with maximum degree of $8$.
\item For $\mcal{C}_{R}^{b}$ we limit ourselves to regular LDGM profiles.
\end{enumerate}

\begin{figure}[htbp]
{\centering
\subfigure[\label{fig_Simulation} BER for $\mbf{b}_{S}$ using design rate of $5.4$bits/sec/Hz with 64QAM.]{\includegraphics[width=3.5in]{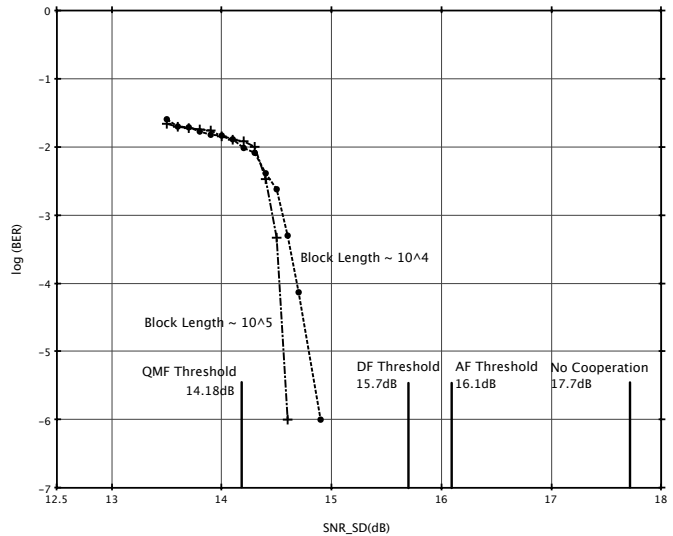}}
\subfigure[\label{fig_SimulationR}BER simulation for $\mbf{b}_{R}$ using design rate of $5.4$bits/sec/Hz with 64QAM.]{\includegraphics[width=3.5in]{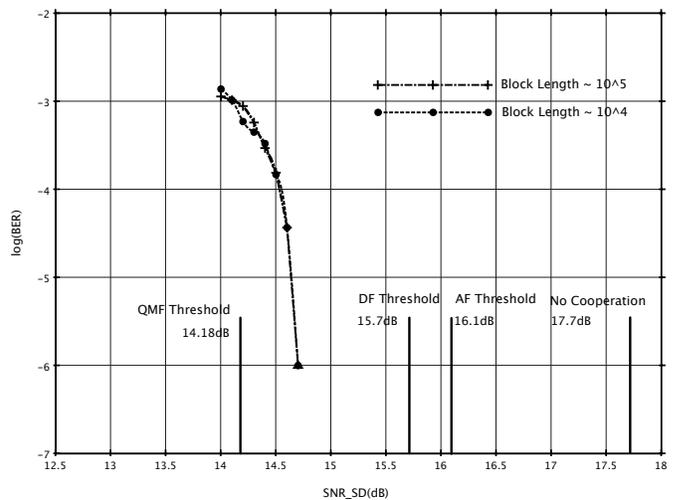}}
\caption{Code design simulation results.}
}
\end{figure}

Using these heuristics we design the following degree profile for the system parameters in this example. 
\begin{align*}
\lambda_{S}(x)&=0.28x+0.32x^{2}+0.28x^{3}+0.12x^{6}+0.0009x^{7}\\
\rho_{S}(x)&=0.04x^{28}+0.96x^{29}\\
\lambda_{R}(x)&=x^{4},
\rho_{R}(x)=x^{9}
\end{align*}
Simulation results for the bit error rate in decoding of $\mbf{b}_{S}$ using codes (with block lengths $\approx10^{4}$ and $\approx10^{5}$) drawn from above profiles are shown in Fig.~\ref{fig_Simulation} using PBICM with $64$QAM modulation, one bit scalar quantizer and an ideal interleaver. As shown the BER performance is $\leq1$dB of the QMF threshold. For the single relay scenario, the information-theoretic thresholds for QMF and CF are identical, therefore as a reference for comparison thresholds for DF, AF, and the no-cooperation case are also shown. The DF and the AF thresholds are computed using the following expressions. Derivations follow standard analysis of the schemes and are omitted here.
\begin{align*}
&\mcal{R}_{\mathrm{DF},n} =\max_{f\in[0,1]} \\&\min\lbp f\C_n\lp\SNR_{SR}\rp , (1-f)\C_n\lp\SNR_{RD}\rp + \C_n\lp\SNR_{SD}\rp\rbp\\
&\mcal{R}_{\mathrm{AF},n} =\frac{1}{2}\C_n\lp\SNR_{SD}\rp + \frac{1}{2}\C_n\lp\SNR_{\rm{eff}}\rp \\
&\SNR_{\rm{eff}} = \SNR_{SD} + \frac{\SNR_{SR}\SNR_{RD}}{1+\SNR_{SR}+\SNR_{RD}}
\end{align*}
The optimal listening time for DF is determined by the channel parameters, while that for AF is always $1/2$.

For the DBLAST architecture, $\mbf{b}_{R}$ must also be reliably decoded at or below the target SNR (for successive interference cancellation to work). Fig.~\ref{fig_SimulationR} shows the BER for $\mbf{b}_{R}$ which is also within $\leq1$dB of the QMF threshold for both of the block-lengths.

\section{Conclusions}
\label{sec_discussion}
The QMF relaying scheme has the following key advantages over other known relaying schemes such as AF, DF, and CF. 
\begin{enumerate}
\item For the single relay network, it outperforms AF and DF at high SNR.
\item For the single relay network, it achieves the same performance as CF but reduces channel feedback overhead. Unlike CF, QMF does not require knowledge of forward channel strength at the relay.
\item For arbitrary relay networks with multiple relays, QMF achieves better high SNR performance than AF, DF and CF.
\end{enumerate}
In this paper, a low-complexity channel coding framework is developed for QMF relaying. For the single relay network, the framework performs within $(0.5-1)$dB of fundamental limits. 

The techniques presented here can be extended to complex system scenarios, which are discussed below.

\subsection{Multiple Relays}
When there is more than one relay in the system, the proposed factor graph extends in a straightforward manner.
Optimal listening schedules can be computed for each of the relays. As proposed, the source would use an LDPC code and each relay would use an LDGM code based on its respective schedule. The joint factor graph would include multiple LDGM sub-graphs. 

The DBLAST architecture proposed in this paper extends naturally to networks with one level of multiple non-interfering relays e.g. the diamond network. As discussed previously, DBLAST significantly reduces the complexity of the factor graph. 
DBLAST requires that all codewords from relays are decoded correctly at destination in order to permit successive interference cancellation. This additional constraint does not lead to a reduction in the QMF information-theoretic achievable rate. In fact, such a requirement is explicitly considered in the probability of error analysis for the QMF scheme in \cite{2010arXiv1002.3188L}.

However, some challenges for multiple relay networks remain to be addressed. When the relays can hear one another or the source can reach the destination via \emph{multiple hops}, it is unclear how the DBLAST architecture can be applied. In such scenarios, an alternate space-time architecture must be considered. Moreover as the number of relays increase, the channel knowledge overhead required to compute optimal listening schedules becomes large. Practical techniques at the physical and MAC layers are required to address this complexity. These are considered as directions for future work. 




\subsection{Rate Adaptation and Hybrid ARQ}
In the link design example, suitable coding rates, constellation and listening fraction are computed for a given set of operating channel conditions. 
However, optimizing codes based on instantaneous channel conditions is not feasible in practice. Under commonly used rate adaptation mechanisms, terminals switch between a few candidate codes and a few candidate constellations based on channel conditions. Cooperative links need to consider multiple channel parameters to determine transmission rates i.e. for a single relay three SNR parameters are required  as opposed to just one for a point-to-point link. This makes rate adaptation schedules for relay networks more complex. 
An advantage of QMF relaying is that rate adaptation schedules depend only on the ability of the destination to decode as opposed to DF, where adaptation must consider decoding at relays as well. 

Modern adaptation mechanisms like hybrid automatic repeat request (HARQ) can be incorporated into the proposed framework. Additional parity bits for refinement sent from the source after receiving a repeat request from the destination. It  can be cooperatively delivered to the destination using QMF relaying. The joint decoding factor graph is expanded to incorporate these refinement parity bits and the decoding algorithm remains unchanged.

\section*{Acknowledgements}
The authors acknowledge Prof. R\"{u}diger Urbanke for fruitful discussions leading to the choice of LDPC-LDGM structures. We also acknowledge the students, faculty and sponsors of the Berkeley Wireless Research Center and support of the Center for Circuit \& System Solutions (C2S2) Focus Center, one of six research centers funded under the Focus Center Research Program, a Semiconductor Research Corporation program.

\appendix
\section{Appendix: DBLAST Space-Time Architecture}
\label{sec_dblast}

The QMF relaying scheme introduces correlation between $\mbf{x}_{S}$ and $\mbf{x}_{R}$, which can be thought of as coding across transmit antennas in a MIMO channel.  A natural space-time architecture for such a channel is DBLAST. Using DBLAST for the relay channel has also been proposed in \cite{4303346}\cite{1532486}\cite{BLTJ:BLTJ2015}\cite{1600070}. It relies on introducing a delay of one block at the relay and using successive interference cancellation (SIC) at the destination. At the $k$-th block the destination receives the superposition of the following:
\begin{itemize}
\item signal from the source containing the codeword sent at block $k$, namely, $\mb{x}_{S}(m_k)$
\item signal from the relay containing the side information about the source's codeword at block $k-1$, namely, $\mb{x}_{R}(q_{k-1})$
\end{itemize} 
Messages sent from the source are independent across blocks.  At the $k$-th block, the destination jointly decodes block $k-1$ (message $m_{k-1}$ and side information $\mb{x}_{R}(q_{k-1})$) by treating $\mb{x}_{S}(m_k)$ as Gaussian noise. The receiver subtracts relay's codeword $\mb{x}_{R}(q_{k-1})$ from its received signal $\mb{Y}[k]$ and keeps the residual $\mb{\wtild{Y}}[k]$ for decoding the next block. This architecture allows the use of a simplified equivalent channel model. Note that the one-block delay introduced at $R$ has the added benefit of allowing time for QMF processing at $R$. 

\subsubsection{Simplified Channel Model}
The equivalent channel model is shown in Fig.~\ref{fig_EquivCh}. For decoding the block $k-1$ message $m_{k-1}$, the decoder takes two inputs $\mb{Y}[k]$ and $\mb{\wtild{Y}}[k-1]$. We can think of $\mb{Y}[k]$ and $\mb{\wtild{Y}}[k-1]$ as two orthogonal links with independent Gaussian noise. Therefore, for the purpose of code design we can alternatively investigate a simpler model depicted in Figure \ref{fig_EquivCh}. In this model,
\begin{align*}
\mbf{Y}_{ij} &= \mbf{h}_{ij}\mbf{x}_{i} + \mbf{Z}_{ij},\ (i,j)=(R,D), (S,D), \\
\mbf{y}_{SR} &= h_{SR}\mbf{x}_{S} + \mbf{z}_{SR}
\end{align*}
As an example, let us consider a scenario where $D$ has two receive antennas $(M=2)$. In that case, the DBLAST equivalent channel becomes \cite{1197843}:
\begin{align*}
\mbf{y}_{ij} &= h_{ij}\mbf{x}_{i} + \mbf{z}_{ij},\ (i,j)=(R,D), (S,D), \\
\intertext{where,}
h_{SD} &= ||\mbf{h}_1||, \ h_{RD} = \sqrt{||\mbf{h}_{2\perp 1}||^2 + \frac{||\mbf{h}_{2\parallel 1}||^2}{1+P_S ||\mbf{h}_1||^2}}
\end{align*}
$\mbf{h}_{2\perp 1}$ and $\mbf{h}_{2\parallel 1}$ denote the perpendicular and parallel components of $\mbf{h}_{2}$ with respect to $\mbf{h}_{1}$, respectively. The signal-to-noise ratios of the three links are $\SNR_{SR}=|h_{SR}|^{2}P_{S}$, $\SNR_{SD}=|h_{SD}|^{2}P_{S}$, and $\SNR_{RD}=|h_{RD}|^{2}P_{R}$ respectively. 

\begin{remark}
Consider the original channel and the DBLAST-equivalent channel. 
Note that the capacities of these two channels are within two bits of each other. This is based on the following observations:
\begin{IEEEenumerate}
\item \textit{The min-cut upper bound for both channels are within one bit of each other (for any listening fraction $f\in[0,1]$).} 
\par The mutual information across cut $\{S\},\{R,D\}$ remains unchanged between the two channels. Consider the mutual information across the cut $\{S,R\},\{D\}$. It is known that SIC achieves the sum capacity of multiple-access channels. In the original channel (Fig.~\ref{fig_HDR}) $S$ and $R$ have unlimited cooperation. As a result, the min-cut bound for DBLAST incurs a power-gain loss of at most $(1-f)$ bits. 
\item \textit{QMF relaying scheme achieves the min-cut upper bound to within one bit for the two channels} \cite{2009arXiv0906.5394A}.
\end{IEEEenumerate}
\end{remark}

\bibliographystyle{IEEEtran} 
\bibliography{IEEEabrv,references} 

\begin{thebibliography}{10}
\providecommand{\url}[1]{#1}
\csname url@samestyle\endcsname
\providecommand{\newblock}{\relax}
\providecommand{\bibinfo}[2]{#2}
\providecommand{\BIBentrySTDinterwordspacing}{\spaceskip=0pt\relax}
\providecommand{\BIBentryALTinterwordstretchfactor}{4}
\providecommand{\BIBentryALTinterwordspacing}{\spaceskip=\fontdimen2\font plus
\BIBentryALTinterwordstretchfactor\fontdimen3\font minus
  \fontdimen4\font\relax}
\providecommand{\BIBforeignlanguage}[2]{{%
\expandafter\ifx\csname l@#1\endcsname\relax
\typeout{** WARNING: IEEEtran.bst: No hyphenation pattern has been}%
\typeout{** loaded for the language `#1'. Using the pattern for}%
\typeout{** the default language instead.}%
\else
\language=\csname l@#1\endcsname
\fi
#2}}
\providecommand{\BIBdecl}{\relax}
\BIBdecl

\bibitem{2009arXiv0906.5394A}
A.~Avestimehr, S.~Diggavi, and D.~Tse, ``Wireless network information flow: A
  deterministic approach,'' \emph{Information Theory, IEEE Transactions on},
  vol.~57, no.~4, pp. 1872 --1905, april 2011.

\bibitem{5205885}
V.~Nagpal, S.~Pawar, D.~Tse, and B.~Nikolic, ``Cooperative multiplexing in the
  multiple antenna half duplex relay channel,'' in \emph{Information Theory,
  2009. ISIT 2009. IEEE International Symposium on}, june 2009, pp. 1438
  --1442.

\bibitem{1056084}
T.~Cover and A.~Gamal, ``Capacity theorems for the relay channel,''
  \emph{Information Theory, IEEE Transactions on}, vol.~25, no.~5, pp. 572 --
  584, sep 1979.

\bibitem{1362898}
J.~Laneman, D.~Tse, and G.~Wornell, ``Cooperative diversity in wireless
  networks: Efficient protocols and outage behavior,'' \emph{Information
  Theory, IEEE Transactions on}, vol.~50, no.~12, pp. 3062 -- 3080, dec. 2004.

\bibitem{4797531}
S.~Pawar, A.~Avestimehr, and D.~Tse, ``Diversity-multiplexing tradeoff of the
  half-duplex relay channel,'' in \emph{Communication, Control, and Computing,
  2008 46th Annual Allerton Conference on}, 23-26 2008, pp. 27 --33.

\bibitem{2010arXiv1002.3188L}
S.~H. Lim, Y.-H. Kim, A.~El~Gamal, and S.-Y. Chung, ``Noisy network coding,''
  \emph{Information Theory, IEEE Transactions on}, vol.~57, no.~5, pp. 3132
  --3152, may 2011.

\bibitem{WangTse_09}
I.-H. Wang and D.~Tse, ``Interference mitigation through limited receiver
  cooperation,'' \emph{Information Theory, IEEE Transactions on}, vol.~57,
  no.~5, pp. 2913 --2940, may 2011.

\bibitem{1204784}
B.~Zhao and M.~Valenti, ``Distributed turbo coded diversity for relay
  channel,'' \emph{Electronics Letters}, vol.~39, no.~10, pp. 786 -- 787, may
  2003.

\bibitem{1023492}
T.~Hunter and A.~Nosratinia, ``Cooperation diversity through coding,'' in
  \emph{Information Theory, 2002. Proceedings. 2002 IEEE International
  Symposium on}, 2002, p. 220.

\bibitem{1261324}
M.~Janani, A.~Hedayat, T.~Hunter, and A.~Nosratinia, ``Coded cooperation in
  wireless communications: space-time transmission and iterative decoding,''
  \emph{Signal Processing, IEEE Transactions on}, vol.~52, no.~2, pp. 362 --
  371, feb. 2004.

\bibitem{1532486}
Z.~Zhang and T.~Duman, ``Capacity-approaching turbo coding and iterative
  decoding for relay channels,'' \emph{Communications, IEEE Transactions on},
  vol.~53, no.~11, pp. 1895 -- 1905, nov. 2005.

\bibitem{4303346}
Z.~Zhang and T.~M. Duman, ``Capacity approaching turbo coding for half-duplex
  relaying,'' \emph{Communications, IEEE Transactions on}, vol.~55, no.~9, p.
  1822, sept. 2007.

\bibitem{4107948}
A.~Chakrabarti, A.~D. Baynast, A.~Sabharwal, and B.~Aazhang, ``Low density
  parity check codes for the relay channel,'' \emph{Selected Areas in
  Communications, IEEE Journal on}, vol.~25, no.~2, pp. 280 --291, february
  2007.

\bibitem{4305411}
P.~Razaghi and W.~Yu, ``Bilayer low-density parity-check codes for
  decode-and-forward in relay channels,'' \emph{Information Theory, IEEE
  Transactions on}, vol.~53, no.~10, pp. 3723 --3739, oct. 2007.

\bibitem{5513451}
T.~V. Nguyen, A.~Nosratinia, and D.~Divsalar, ``Bilayer protograph codes for
  half-duplex relay channels,'' in \emph{Information Theory Proceedings (ISIT),
  2010 IEEE International Symposium on}, june 2010, pp. 948 --952.

\bibitem{4259765}
P.~Razaghi, M.~Aleksic, and W.~Yu, ``Bit-interleaved coded modulation for the
  relay channel using bilayer ldpc codes,'' in \emph{Information Theory, 2007.
  CWIT '07. 10th Canadian Workshop on}, 6-8 2007, pp. 101 --104.

\bibitem{1721040}
M.~Uppal, Z.~Liu, V.~Stankovic, and Z.~Xiong, ``Compress-forward coding with
  bpsk modulation for the half-duplex gaussian relay channel,'' \emph{Trans.
  Sig. Proc.}, vol.~57, no.~11, pp. 4467--4481, 2009.

\bibitem{5604326}
M.~Uppal, G.~Yue, X.~Wang, and Z.~Xiong, ``A rateless coded protocol for
  half-duplex wireless relay channels,'' \emph{Signal Processing, IEEE
  Transactions on}, vol.~59, no.~1, pp. 209 --222, jan. 2011.

\bibitem{2010arXiv1005.1284O}
A.~Ozgur and S.~Diggavi, ``Approximately achieving gaussian relay network
  capacity with lattice codes,'' \emph{ArXiv e-prints}, May 2010.

\bibitem{669123}
G.~Caire, G.~Taricco, and E.~Biglieri, ``Bit-interleaved coded modulation,''
  \emph{Information Theory, IEEE Transactions on}, vol.~44, no.~3, pp. 927
  --946, may 1998.

\bibitem{910578}
T.~Richardson, M.~Shokrollahi, and R.~Urbanke, ``Design of capacity-approaching
  irregular low-density parity-check codes,'' \emph{Information Theory, IEEE
  Transactions on}, vol.~47, no.~2, pp. 619 --637, feb 2001.

\bibitem{910577}
T.~Richardson and R.~Urbanke, ``The capacity of low-density parity-check codes
  under message-passing decoding,'' \emph{Information Theory, IEEE Transactions
  on}, vol.~47, no.~2, pp. 599 --618, feb 2001.

\bibitem{4698542}
Y.~Fan, H.~Poor, and J.~Thompson, ``Cooperative multiplexing in full-duplex
  multi-antenna relay networks,'' in \emph{Global Telecommunications
  Conference, 2008. IEEE GLOBECOM 2008. IEEE}, 30 2008-dec. 4 2008, pp. 1 --5.

\bibitem{5706940}
V.~Nagpal, I.-H. Wang, M.~Jorgovanovic, D.~Tse, and B.~Nikoli{\'c},
  ``Quantize-map-and-forward relaying: Coding and system design,'' in
  \emph{Communication, Control, and Computing (Allerton), 2010 48th Annual
  Allerton Conference on}, 29 2010-oct. 1 2010, pp. 443 --450.

\bibitem{910572}
F.~Kschischang, B.~Frey, and H.-A. Loeliger, ``Factor graphs and the
  sum-product algorithm,'' \emph{Information Theory, IEEE Transactions on},
  vol.~47, no.~2, pp. 498 --519, feb 2001.

\bibitem{825794}
S.~Aji and R.~McEliece, ``The generalized distributive law,'' \emph{Information
  Theory, IEEE Transactions on}, vol.~46, no.~2, pp. 325 --343, mar 2000.

\bibitem{DBLP:journals/corr/abs-1008-1766}
A.~Bennatan, S.~Shamai, and A.~R. Calderbank, ``In praise of bad codes for
  multi-terminal communications,'' \emph{CoRR}, vol. abs/1008.1766, 2010.

\bibitem{DBLP:journals/corr/cs-IT-0408008}
E.~Martinian and J.~S. Yedidia, ``Iterative quantization using codes on
  graphs,'' \emph{CoRR}, vol. cs.IT/0408008, 2004.

\bibitem{BLTJ:BLTJ2015}
\BIBentryALTinterwordspacing
G.~J. Foschini, ``Layered space-time architecture for wireless communication in
  a fading environment when using multi-element antennas,'' \emph{Bell Labs
  Technical Journal}, vol.~1, no.~2, pp. 41--59, 1996. [Online]. Available:
  \url{http://dx.doi.org/10.1002/bltj.2015}
\BIBentrySTDinterwordspacing

\bibitem{1600070}
G.~Kramer, ``Distributed and layered codes for relaying,'' in \emph{Signals,
  Systems and Computers, 2005. Conference Record of the Thirty-Ninth Asilomar
  Conference on}, october 2005, pp. 1752 -- 1756.

\bibitem{4313147}
G.~Kraidy, N.~Gresset, and J.~Boutros, ``Coding for the non-orthogonal
  amplify-and-forward cooperative channel,'' in \emph{Information Theory
  Workshop, 2007. ITW '07. IEEE}, 2-6 2007, pp. 626 --631.

\bibitem{4939335}
M.~Benjillali and L.~Szczecinski, ``A simple detect-and-forward scheme in
  fading channels,'' \emph{Communications Letters, IEEE}, vol.~13, no.~5, pp.
  309 --311, may 2009.

\bibitem{HouSiegel_03}
J.~Hou, P.~H. Siegel, L.~B. Milstein, and H.~D. Pfister, ``Capacity-approaching
  bandwidth-efficient coded modulation schemes based on low-density
  parity-check codes,'' \emph{Information Theory, IEEE Transactions on},
  vol.~49, no.~9, pp. 2141--2155, 2003.

\bibitem{IngberFeder_10}
A.~Ingber and M.~Feder, ``Parallel bit interleaved coded modulation,''
  \emph{Proceedings of Annual Allerton Conference on Communications, Control,
  and Computing}, September 2010.

\bibitem{4305423}
M.~Yuksel and E.~Erkip, ``Multiple-antenna cooperative wireless systems: A
  diversity-multiplexing tradeoff perspective,'' \emph{Information Theory, IEEE
  Transactions on}, vol.~53, no.~10, pp. 3371 --3393, oct. 2007.

\bibitem{910580}
S.-Y. Chung, T.~Richardson, and R.~Urbanke, ``Analysis of sum-product decoding
  of low-density parity-check codes using a gaussian approximation,''
  \emph{Information Theory, IEEE Transactions on}, vol.~47, no.~2, pp. 657
  --670, feb 2001.

\bibitem{1197843}
L.~Zheng and D.~Tse, ``Diversity and multiplexing: a fundamental tradeoff in
  multiple-antenna channels,'' \emph{Information Theory, IEEE Transactions on},
  vol.~49, no.~5, pp. 1073 -- 1096, may 2003.

\end{thebibliography}

\begin{IEEEbiography}[{\includegraphics[width=1in,height=1.25in,clip,keepaspectratio] {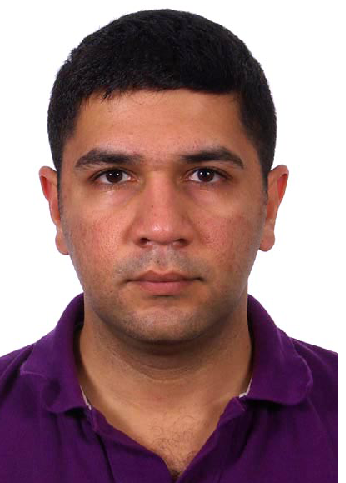} }]{Vinayak Nagpal}
received the B. Engg. degree from University of Pune, India in 2003 and M.S. from Chalmers University of Technology, Sweden in 2006. He received the Ph.D. degree from University of California at Berkeley, USA in 2012 under the guidance of Prof. Borivoje Nikoli\'{c}. Since then he is affiliated with Nokia Research Center, Berkeley, USA. Previously he has held positions at Conexant Systems, Pune India (2003), National Radio Astronomy Observatory, Charlottesville, VA (2005), and Harvard Smithsonian Center for Astrophysics, Cambridge MA (2006).
His research interests include wireless networks and real time signal processing. 
\end{IEEEbiography}

\begin{IEEEbiography}
[{\includegraphics[width=1in,height=1.25in,clip,keepaspectratio] {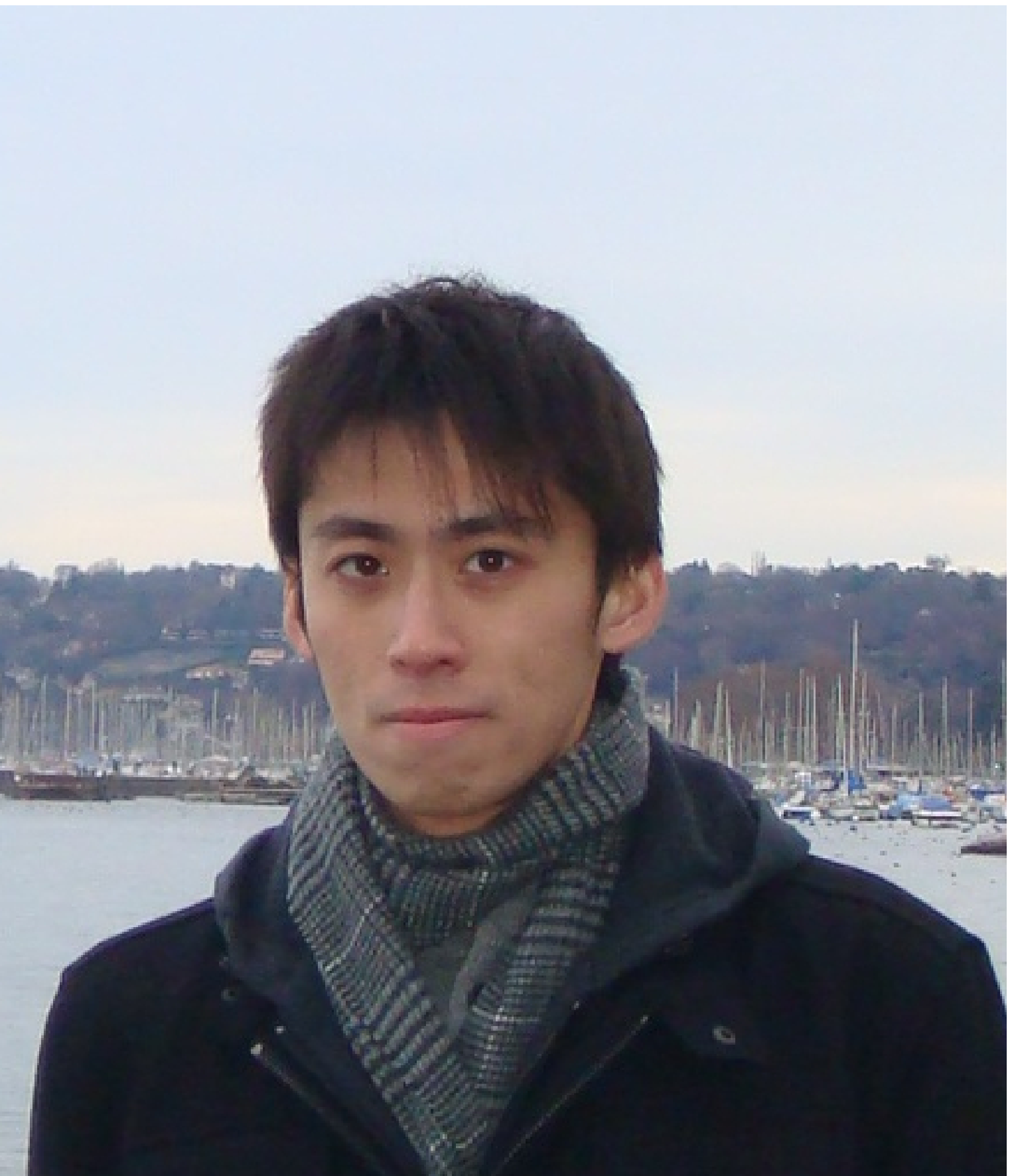} }]{I-Hsiang Wang}
received the B.S. degree in Electrical Engineering from National Taiwan University, Taiwan in 2006. He received a Ph.D. degree in electrical engineering and computer
sciences from University of California at Berkeley, USA, in 2011. Since
2011, he has been affiliated with \'{E}cole Polytechnique F\'{e}d\'{e}rale de Lausanne,
Switzerland, as a postdoctoral researcher. His research interests include network
information theory, wireless networks, coding theory and network coding. Dr.
Wang received a 2-year Vodafone Graduate Fellowship in 2006.
\end{IEEEbiography}

\begin{IEEEbiography}
[{\includegraphics[width=1in,height=1.25in,clip,keepaspectratio] {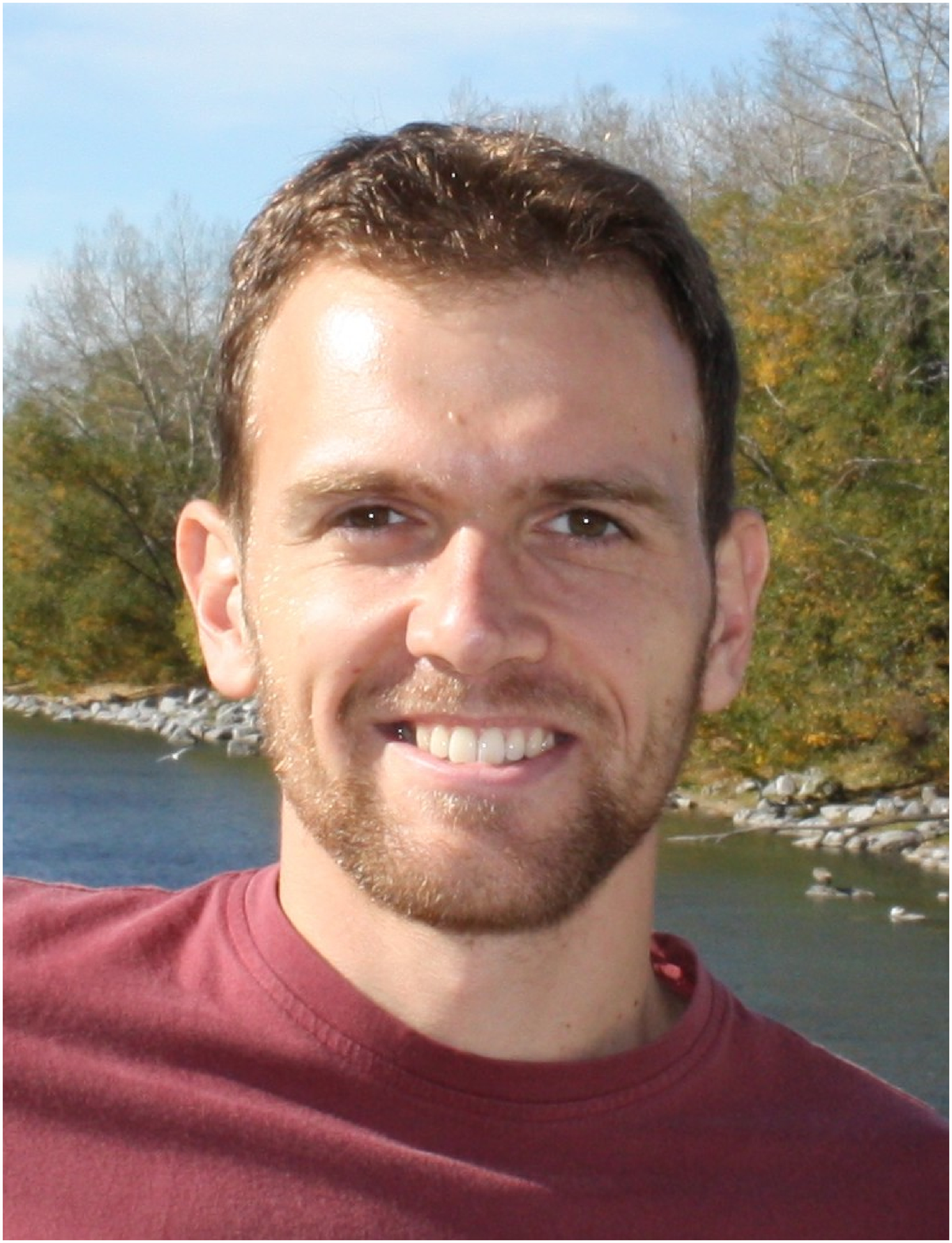} }]{Milos Jorgovanovic}
received his Dipl. Ing. degree in Electrical Engineering from University of Belgrade, Serbia in 2007 and M.Sc. degree from University of California at Berkeley in 2010. He is currently working towards his Ph.D. degree at University of California at Berkeley under guidance of Prof. Borivoje Nikoli\'{c}. 
He held internship positions with Kodak European Research Center in Cambridge, UK (2006), Technical University of Berlin, Germany (2009) and Samsung Mobile in Richardson, TX (2010). His research interests include MIMO detection algorithms and architectures, wireless communication systems design, signal processing for digital communications and digital integrated circuit design.
\end{IEEEbiography}

\begin{IEEEbiography}[{\includegraphics[width=1in,height=1.25in,clip,keepaspectratio] {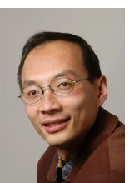} }]{David Tse} 
received the B.A.Sc. degree in systems design engineering from the University of Waterloo, Waterloo, ON, Canada, in 1989, and the M.S. and Ph.D. degrees in electrical engineering from the Massachusetts Institute of Technology, Cambridge, in 1991 and 1994, respectively.
From 1994 to 1995, he was a Postdoctoral Member of Technical Staff at
AT\&T Bell Laboratories. Since 1995, he has been with the Department of Electrical
Engineering and Computer Sciences, University of California at Berkeley,
where he is currently a Professor.
Dr. Tse received a 1967 NSERC 4-year graduate fellowship from the government
of Canada in 1989, a NSF CAREER award in 1998, the Best Paper Awards
at the Infocom 1998 and Infocom 2001 conferences, the Erlang Prize in 2000
from the INFORMS Applied Probability Society, the IEEE Communications
and Information Theory Society Joint Paper Award in 2001, the Information
Theory Society Paper Award in 2003, and the 2009 Frederick Emmons Terman
Award from the American Society for Engineering Education. He has given plenary
talks at international conferences such as ICASSP in 2006, MobiCom in
2007, CISS in 2008, and ISIT in 2009. He was the Technical Program Cochair of
the International Symposium on Information Theory in 2004 and was an Associate
Editor of the IEEE Transactions on Information theory from 2001 to
2003. He is a coauthor, with P. Viswanath, of the text Fundamentals of Wireless
Communication, which has been used in over 60 institutions around the world.
\end{IEEEbiography}

\begin{IEEEbiography}[{\includegraphics[width=1in,height=1.25in,clip,keepaspectratio] {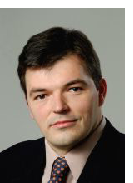} }]{Borivoje Nikoli\'{c}} 
received the Dipl.Ing. and M.Sc. degrees in electrical engineering from the University of Belgrade, Serbia, in 1992 and 1994, respectively, and the Ph.D. degree from the University of California at Davis in 1999.
He lectured electronics courses at the University of Belgrade from 1992 to 1996. He spent two years with Silicon Systems, Inc., Texas Instruments Storage Products Group, San Jose, CA, working on disk-drive signal processing electronics. In 1999, he joined the Department of Electrical Engineering and
Computer Sciences, University of California at Berkeley, where he is now a Professor. His research activities include digital and analog integrated circuit design and VLSI implementation of communications and signal processing algorithms. He is a co-author of the book Digital Integrated Circuits: A Design Perspective (2nd ed., Prentice-Hall, 2003).
Dr. Nikoli\'{c} received the NSF CAREER award in 2003, College of Engineering Best Doctoral Dissertation Prize and Anil K. Jain Prize for the Best Doctoral Dissertation in Electrical and Computer Engineering at University of California at Davis in 1999, as well as the City of Belgrade Award for the Best Diploma Thesis in 1992. For work with his students and colleagues he has received the best paper awards at the IEEE International Solid-State Circuits Conference, Symposium on VLSI Circuits, IEEE International SOI Conference and the ACM/IEEE International Symposium of Low-Power Electronics.
\end{IEEEbiography}

\end{document}